\documentclass[12pt]{article}
\usepackage[utf8]{inputenc}
\usepackage[english]{babel}
\usepackage[centertags]{amsmath}
\usepackage{amssymb}
\usepackage{bm}
\usepackage{amsbsy}
\usepackage{graphicx}
\usepackage{appendix}
\usepackage{mathrsfs}
\usepackage{epsfig}
\usepackage{color}
\usepackage{euscript}
\usepackage{ulem}
\usepackage{multirow}
\usepackage{cite}

\newcommand{\arctanh}{\mathop{\mathrm{arctanh}}\nolimits}

\definecolor{dgreen}{rgb}{0,0.6,0}

\definecolor{darkblue}{rgb}{0., 0, 1}

\definecolor{purple}{rgb}{0.65,0.,0.78}

\definecolor{orange}{rgb}{0.89,0.3,0.12}

\usepackage{jheppubm}

\newcommand{\nn}{\nonumber}
\newcommand{\be}{\begin{equation}}
\newcommand{\ee}{\end{equation}}
\newcommand{\bea}{\begin{eqnarray}}
\newcommand{\eea}{\end{eqnarray}}
\newcommand{\arcsinh}{\mbox{arcsinh}}

\newcommand{\fb}{\mathfrak{b}}
\newcommand{\fc}{\mathfrak{c}}
\newcommand{\ff}{\mathfrak{f}}

\newcommand{\fg}{\mathfrak{g}}

\newcommand{\fz}{\mathfrak{z}}

\newcommand{\fG}{\mathfrak{G}}

\newcommand{\cV}{{\cal V}}
\newcommand{\cZ}{{\cal Z}}
\newcommand{\cX}{{\cal X}}
\newcommand{\cY}{{\cal Y}}
\newcommand{\cH}{{\cal H}}

\numberwithin{equation}{section}

\title{
Beta-Functions and RG flows for Holographic QCD with   Heavy and Light Quarks: Isotropic case
}

\author{Irina Ya. Aref'eva$^a$, Ali Hajilou$^a$, Pavel Slepov$^a$ and  Marina Usova$^a$}

\affiliation{$^a$Steklov Mathematical Institute, Russian Academy of  Sciences, \\ Gubkina str. 8, 119991, Moscow, Russia}

\emailAdd{arefeva@mi-ras.ru}
\emailAdd{hajilou@mi-ras.ru}
\emailAdd{slepov@mi-ras.ru}
\emailAdd{usovamk@mi-ras.ru}

 \abstract{In a previous paper \cite{AHSU}, we investigated the dependence of the running coupling constant on temperature and chemical potential for holographic models of the light and heavy quarks, supported by an Einstein-dilaton-Maxwell action. In this paper, we study the dependence of the corresponding $\beta$-functions on the temperature and the chemical potential. As in the previous paper, we give special attention to the behavior of the $\beta$-functions near the 1st order phase transitions.

We consider different types of boundary conditions for the dilaton. Only one of the possible boundary conditions yields results that agree with lattice calculations at zero chemical potential. The corresponding $\beta$-functions are negative and exhibit jumps at the 1st order phase transitions. We also show that the RG fluxes are invariant with respect to the choice of the boundary conditions and that our exact solutions for the light and heavy quarks are unstable, as expected, given their negative dilaton potentials.
}


\begin{document}

\maketitle

\newpage

\section{Introduction}\label{Intro}
The renormalization group (RG) is an approach that refers to changing the physical system as we considered different scales of energy \cite{BogSchirkov,Wilson:1973jj,ModernTextBook}.
In fact, the RG approach provides a systematic picture for describing a physical system with many physical parameters. The dependence of the coupling constants of the physical system on the energy scale can be described by the $\beta$-function of the theory \cite{GellMannLow,{Callan:1970yg,Symanzik:1970rt,Wilson:1973jj}}. The RG flows are a variation of the fields versus the energy scale of the theory. 
The RG is widely used in various areas of the modern theoretical physics \cite{Zinn,Vasiliev}.
\\

Holographic duality describes a correspondence between a class of strongly coupled field theories and weakly-coupled gravitational theories \cite{Maldacena:1997re}. 
The strongly coupled regime of gauge theories can be explored using holography, this duality gives an important result on the shear viscosity, which agrees with experimental results \cite{Policastro:2001yc,Kovtun:2004de}, for a review of other phenomenological applications see refs \cite{Casalderrey-Solana:2011dxg,DeWolfe:2013cua,Arefeva:2014kyw,Ammon:2015wua}.
The holographic RG flow sets up an explanation of the RG flow in terms of a gravitational model coupled to the dilaton field \cite{deBoer:1999tgo,Boonstra:1998mp,Heemskerk:2010hk,Faulkner:2010jy,Lee:2013dln,Bianchi:2001kw,Skenderis:2002wp,Gursoy:2007cb, Gursoy:2007er,Kiritsis:2014kua,Kiritsis:2016kog,Gursoy:2018umf,Ghosh:2017big}. 
The RG flow in this approach has a geometric description and is dual to the gravitational solution with specific asymptotic characteristics in such a way that a holographic coordinate $z$ corresponds to the energy scale of the gauge theory.
Therefore, a Hamiltonian evolution in the holographic direction corresponds to the evolution of the physical system when the scale of the RG changes.
\\

Studies of the holographic RG flow have been widely carried out in \cite{Gursoy:2007cb, Gursoy:2007er,Kiritsis:2014kua,Kiritsis:2016kog,Gursoy:2018umf,Ghosh:2017big} in the context of QCD applications.
The  RG flows for physical systems, including systems with non-zero chemical potential and anisotropic quark-gluon plasma (QGP), are  investigated in \cite{Arefeva:2018hyo,Arefeva:2020aan,Arefeva:2019qen}.
Of particular interest are the exact holographic RG flows that exist for both higher-dimensional \cite{Arefeva:2018jyu} and low-dimensional cases
\cite{Berg:2001ty,Golubtsova:2022hfk,Arkhipova:2024iem,Golubtsova:2024dad}.  
\\

In QFT, the coupling constant can be found as a solution to the RG equation which is governed by the $\beta$-function.
The $\beta$-function encodes the dependence of the coupling constant $\alpha$ on the energy scale of the physical system and is defined as
\be
\beta_{QFT}(\alpha)=\frac {\partial \alpha(E)}{\partial \ln(E)}~,
\ee 
    here $\alpha=\alpha(E)$ is the running coupling and  $E$ denotes the energy scale in QFT.  The holographic $\beta$-function is defined  by \cite{deBoer:1999tgo, Gursoy:2007cb},

 \be \label{beta-z}
\beta(\alpha)= \,\frac{\dot{\alpha} }{\dot{A}}\,,\qquad A=\log B,
 \ee
where $\alpha=\alpha(z)$ is defined  by  the dilaton field  
\eqref{lambda-phi}, dot means the derivative with respect to the holographic coordinate $z$ (see details in the text, in particular, Sec.\,\ref{sec:beta}), and the function $B(z)$ specifying  the warp factor in a holographic metric  \eqref{metric} plays the role of  the energy scale $E$ in the boundary field theory. In simplest holographic models with $B=1/z$ the holographic coordinate $z$ is related to the energy scale $E$ in the boundary field theory as $z\sim \frac{1}{E}$ \cite{Peet:1998wn,Galow:2009kw,Casalderrey-Solana:2011dxg}.
\\

The holographic $\beta$-function allows us to study theories that are close to the specific chosen one. These theories  have the same dilaton potential but  differ in warp factors, i.e., are described by various metrics. This is important to study the stability of the theory, in particular,  under changing boundary conditions. To study the holographic RG flow for given potential, in our case the potential reconstructed by the phenomenologically accepted model, one introduces a dynamical variable $X=X(\varphi)$ that satisfies  the holographic RG flow equation \cite{Kiritsis:2014kua,Kiritsis:2016kog,Gursoy:2018umf,Ghosh:2017big,Arefeva:2018hyo,Arefeva:2019qen,Arefeva:2020aan}. In the case of non-vacuum solutions for the non-zero temperature and the chemical potential, the RG flow is given  by solving the system of equations. In this paper, we start to elaborate on this system of equations to understand the influence of the phase transition on the behavior of their solutions.
\\

Holographic QCD as a non-perturbative approach is a powerful tool for studying the physics of heavy and light quarks.
To study the behavior of $\beta$-functions in various phases, i.e., the quark confinement phases
and QGP, as well as near critical lines, we use holographic QCD models for the heavy and light quarks \cite{Andreev:2006ct,Arefeva:2016rob,Yang:2015aia,Arefeva:2018hyo,Li:2017tdz,Arefeva:2020byn,Arefeva:2022bhx, Hajilou:2021wmz,Arefeva:2021mag,Asadi:2021nbd}. 
We will then be able to capture features of the phase diagram at the low chemical potentials, as well as predict new features at the finite chemical potentials.
It is important to note that the holographic model that we considered in \cite{AHSU} is based on the bottom-up approach. In this phenomenological framework, we describe certain properties of QCD-like theories by manipulating the Einstein-Hilbert action and introducing a dilaton field or other gauge fields.
In constructing these models, gravity is typically coupled to the dilaton and Maxwell fields. The dilaton field is responsible for the running coupling constant in the real QCD, while the Maxwell field accounts for the chemical potential. \\

Let us remind that one  of the goals of experiments at the Large Hadron Collider (LHC), Relativistic Heavy Ion Collider (RHIC), Nuclotron-based Ion Collider fAcility (NICA), and Facility for Antiproton and Ion Research (FAIR) is to study the QCD phase diagram in the (chemical potential, temperature)-plane. Standard QCD calculations, such as perturbation theory, are not applicable in the strong coupling regime. Therefore, to describe the physics of the strongly interacting quark-gluon plasma (QGP) produced in heavy ion collisions (HIC) at RHIC, LHC, and NICA—as well as in future experiments—a non-perturbative approach is required \cite{Casalderrey-Solana:2011dxg,Policastro:2001yc,Kovtun:2004de,Ammon:2015wua,DeWolfe:2013cua,Arefeva:2014kyw}.
From lattice calculations \cite{Brown:1990ev, Philipsen:2016hkv, Guenther:2020jwe, Aarts:2023vsf} and certain effective phenomenological approaches \cite{Fu:2019hdw, Dumm:2021vop, Du:2024wjm}, it is expected that a structure of the QCD phase diagram depends significantly on the quark masses, differing between heavy and light quarks. For this reason, in \cite{AHSU}, we have thoroughly studied the behavior of the running coupling near the 1st order phase transitions in holographic models that describe heavy and light quarks separately. 
\\

In this paper, we study the holographic $\beta$-function near the phase transition lines for the holographic models considered in \cite{AHSU}, which are well-established models that holographically describe QCD for both light and heavy quarks (see \cite{Arefeva:2022avn, Arefeva:2023jjh}).
In this paper, we study the dependence of the corresponding $\beta$-functions on temperature and chemical potential for both heavy and light quarks. Since the behavior of physical quantities near critical points or phase transition lines are important, as in \cite{AHSU}, we give special attention to the behavior of the $\beta$-functions near the 1st order phase transitions. 
Considering different types of the dilaton boundary conditions, we found that only one of the possible boundary conditions yields results consistent with lattice calculations at the zero chemical potential. Our results show that the corresponding $\beta$-functions are negative and exhibit jumps at the 1st order phase transitions.
\\

The paper is organized as follows. In Sec.\,\ref{sec:prelim},
we present  5-dim holographic models for heavy and light quarks and describe  thermodynamic properties
of these models. For the reader's convenience, we duplicate here some details concerning the holographic background from \cite{AHSU} in this Section. In Sec.\,\ref{sec:beta},  we describe  the $\beta$-function for the light and heavy quarks models. In Sec.\,\ref{sec:RG}, we describe RG flows constant for the light and heavy quarks models. In Sec.\,\ref{sec:concl},  we review our main
results. This work is complemented with: Appendix\,\ref{appendix A} where we describe the reconstruction method and solve equations of motion (EOMs),  Appendix\,\ref{approx1}
where we present approximation coefficients of the potential and the gauge kinetic function.

\newpage

\section{Holographic models for the light and heavy quarks }\label{sec:prelim}
\subsection{Solution and background }
\label{sect:s-b}
We consider the Einstein-Maxwell-dilaton (EMd) system with the action \cite{Li:2017tdz,Yang:2015aia}
\bea
S&=&\frac{1}{16\pi G_5}\int d^5x\sqrt{-\fg} \left[R-\frac{\ff_0(\varphi)}{4}F^2-\frac{1}{2}\partial_{\mu}\varphi\partial^{\mu}\varphi-\cV(\varphi)\right],\label{action}
\eea
where $G_5$ is the 5-dimensional Newtonian constant, $g_{\mu\nu}$ is the metric tensor, $\fg=det\,g_{\mu\nu}$ is the determinant of the metric tensor,
$F_{\mu\nu}$ is the electromagnetic tensor of the gauge Maxwell field $A_{\mu}$, $F_{\mu\nu}=\partial_{\mu}A_{\nu}-\partial_{\nu}A_{\mu}$, $\varphi$ is the dilaton field, $\ff_0(\varphi)$ is the gauge kinetic function associated to the Maxwell field, $\cV(\varphi)$ is the potential of the dilaton field $\varphi$.\\

We propose the ansatz for the metric, dilaton field and Maxwell field as \cite{Li:2017tdz,Yang:2015aia}
\bea\label{metric}
ds^2=B^2(z)\left[-g(z)dt^2+d\Vec{x}^2+\frac{dz^2}{g(z)}\right],\label{metric}\\
 \quad  \varphi=\varphi(z),\quad \label{warp-factor} A_{\mu}=\left(A_t(z)~,\Vec{0},0\right)~,
\eea
where 
\be
\label{warp-factor} B(z)=\frac{L\,e^{A(z)}}{z}
\ee is the warp factor, $g(z)$ is the blackening function, $\Vec{x}=(x_1,x_2,x_3)$ and  $A(z)$ is a scale  factor that has different functionality associated to the light and heavy quarks and $L$ is the AdS radius that we set $L=1$.
All functions  depend on the omitted holographic coordinate $z$   and $V(z)=\cV(\varphi(z))$,  
$V_\varphi(z)=\cV_\varphi(\varphi(z))$ and $f_0(z)=\ff_0(\varphi(z))$. 
\\

Varying the action (\ref{action}) and applying the ansatz (\ref{metric})-(\ref{warp-factor}), we obtain the EOMs (\ref{phi2prime})-(\ref{A2primes}) \cite{Arefeva:2023ter,Arefeva:2020vae,Arefeva:2022avn} and solve them with the following boundary conditions 
\bea
  A_t(0) = \mu, \quad A_t(z_h) = 0, \label{eq:4.24} \\
 \quad  g(0) = 1, \qquad g(z_h) = 0, \label{eq:4.25} 
 \eea
which lead to the analytical solutions of the EOMs that, in general, are the same for the light and heavy quarks \eqref{phiprime}-\eqref{Vsol}.

We consider the dilaton field  with different boundary conditions \cite{Slepov:2021gvl,Arefeva:2019yzy, AHSU, Arefeva:2020byn}
 as $\varphi(z,z_0)$ in such a way that
\be\label{phi-z0}
\varphi(z,z_0)\Big|_{z=z_0}=0 \,.
\ee
By choosing $z_0$, three different types of boundary conditions can be imposed:
\bea \label{bc0}
z_0&=&0,\\
\label{bch}
z_0&=&z_h,\\ \label{bce}
z_0&=&\fz(z_h),
\eea
where $\fz (z_h)$ is a smooth function of $z_h$. The dilaton field with the zero boundary condition \eqref{bc0} is denoted by  $\varphi_0(z)$, i.e. $\varphi(z,0)=\varphi_0(z)$.
The first boundary condition \eqref{bch} is denoted by  $\varphi_{z_h}(z)$, and for the second boundary condition \eqref{bce} we have 
\be
\label{phi-z0-gen}
\varphi_\fz(z)=\varphi(z,\fz(z_h)),
\qquad\mbox{i.e.}\qquad
\varphi(\fz(z_h),\fz(z_h))=0.
\ee

Our choice for the light quarks is
\be\label{phi-fz-LQ}
z_0=\fz_{\,_{LQ}}(z_h)=10 \, e^{(-\frac{z_h}{4})}+0.1\,.\ee
To respect the linear Regge trajectories at $T=\mu=0$ for meson spectrum, we have to choose the gauge kinetic function in the form  \cite{Karch:2006pv,Li:2017tdz,Yang:2015aia}
\bea \label{wfLc}
f_0(z)=e^{-c \, z^2-A(z)},
\eea
where $A(z)$ a  scale factor of the light quarks is
\be\label{wfL}
A(z)=-a\log(bz^2+1)~,
\ee
 and $a$, $b$, and $c$ are parameters fitted with the experimental data as $a = 4.046$, $b = 0.01613$ GeV${}^2$ and $c=0.227$   GeV${}^2$ \cite{Li:2017tdz}.  
The Regge spectrum does not depend on the choice of the boundary condition for $\varphi$.
\\
 
The heavy quarks model \cite{Yang:2015aia,Arefeva:2023jjh} contains another warp factor which is specified by the scale factor $A(z)$ given by
\bea \label{scaleHQ}
A(z)=-\frac{\fc}{3}z^2- p\, z^4~,
\eea
$\fc$ and $p$ are parameters fitted with the experimental data as $\fc= 1.16$ GeV${}^2$ and $p = 0.273$ GeV${}^4$. 
The gauge kinetic function for the heavy  quarks model is chosen in the form 
\bea
 ~~~~f_0(z)=e^{-\fc \, z^2-A(z)},
\eea
 compare with \eqref{wfLc}. To solve the EOMs  for the heavy quarks model, we can impose the same boundary conditions (\ref{eq:4.24}) and (\ref{eq:4.25}) and three types of boundary conditions for the dilaton field corresponding to  (\ref{bc0}), (\ref{bch}) and (\ref{bce}). For the heavy quarks model the second boundary condition can be considered as
\bea \label{bceHQ}
z_0&=& \fz_{\,_{HQ}}(z_h) =e^{(-\frac{z_h}{4})} + 0.1 \,.
\eea

We need to emphasize that the first boundary condition $z_0=z_{h}$ (\ref{bch}) is unphysical in accordance with the behavior of the QCD string tension in the Cornell potential as the function of temperature, i.e. $\sigma(T)$, that obtained via lattice calculations \cite{Cardoso:2011hh}. But the second boundary conditions, namely (\ref{phi-fz-LQ}) and  (\ref{bceHQ}), are physical for the light and heavy quarks, respectively (for more details see \cite{AHSU}). Analytical solutions for the heavy quarks model are functionally given by equations \eqref{spsol}-\eqref{Vsol}, the  only difference is the scale factor $A(z)$ and the constant parameter $c$ should be replaced by $\fc$ as a new constant parameter for the heavy quarks model.

\subsection{Phase structure} \label{phaseLQ}

The temperature and entropy for the metric \eqref{metric} can be written as:
\begin{gather}
  \begin{split}
    T &= \cfrac{|g'|}{4 \pi} \, \Bigl|_{z=z_h}
  \end{split}\, , \quad  s = \frac{B^{3}(z_h)}{4}\, \label{eq:2.03}
\end{gather}
here we set $G_5=1$. To get the 1st order  phase transition line we
need to calculate free energy as a function of the  temperature:
\begin{gather}
  F =  \int_{z_h}^{z_{h_2}} s \, T' dz, \label{eq:2.05}
\end{gather}
where $z_{h_2}$ is a second horizon of black hole appearing at $T = 0$, see \cite{Arefeva:2020vae}.  To respect a null energy condition (NEC) we should consider the holographic model for $z_h < z_{h_2}$. \\

\begin{figure}[t!]
  \centering
\includegraphics[scale=0.47]{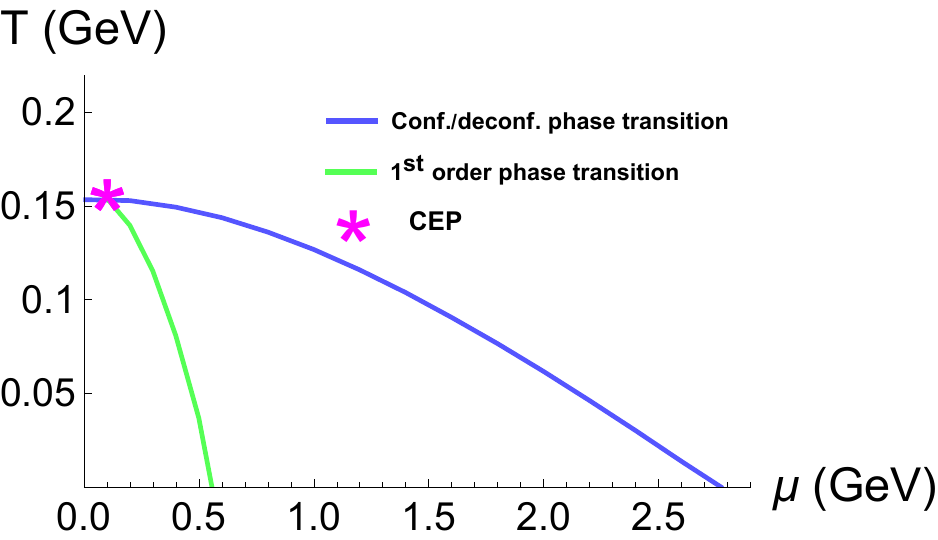} 
\quad\includegraphics[scale=0.52]{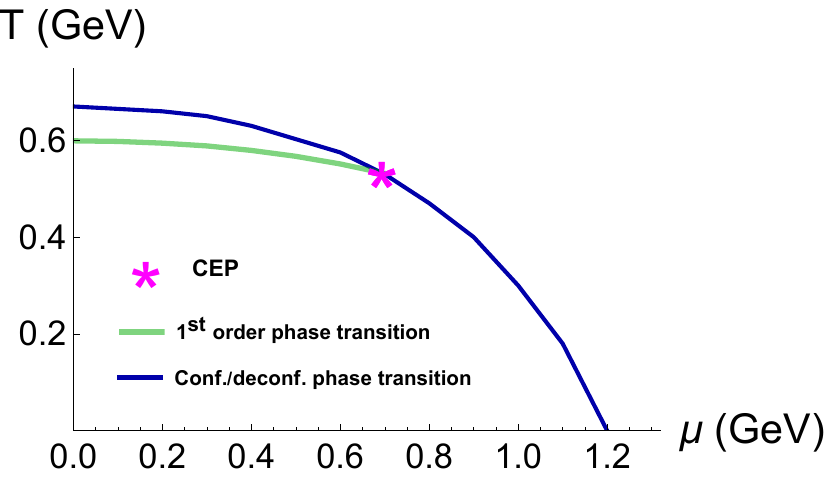}
\\
  A\hspace{200pt}B\\
\caption{Phase structure for the light (A) and heavy quarks model (B). The magenta stars show the critical end points (CEPs). 
  }
 \label{Fig:PD-LH-HQ-paint}
\end{figure}

\begin{figure}[h!]
  \centering
\includegraphics[scale=0.4]{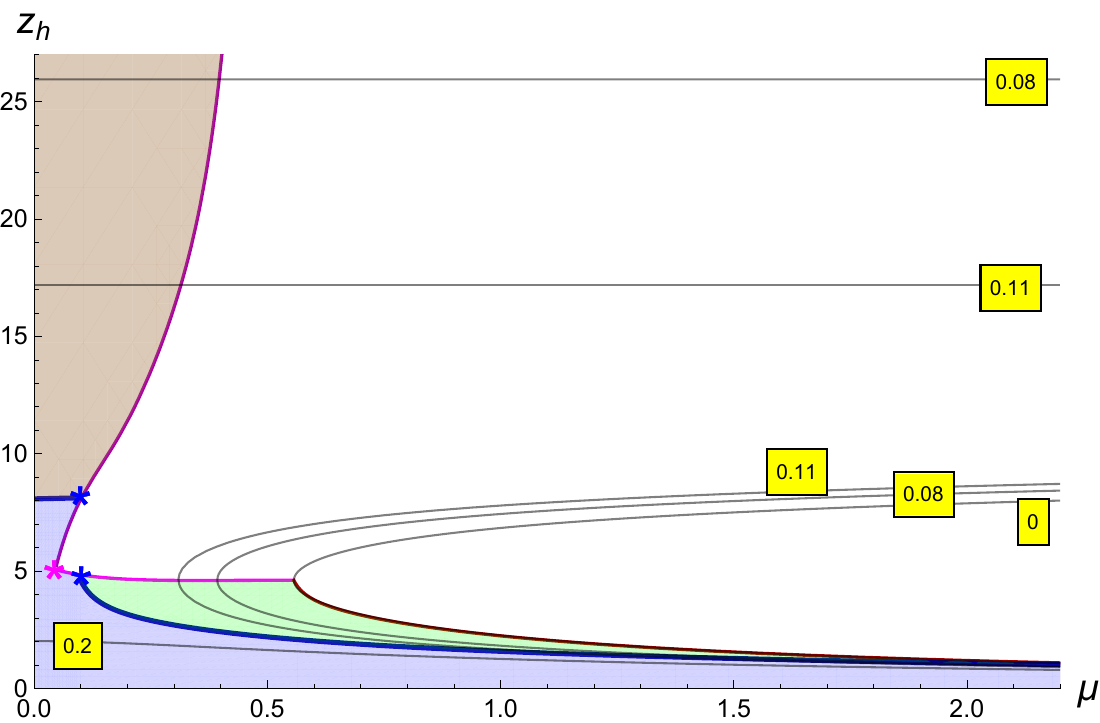} 
\includegraphics[scale=0.29]{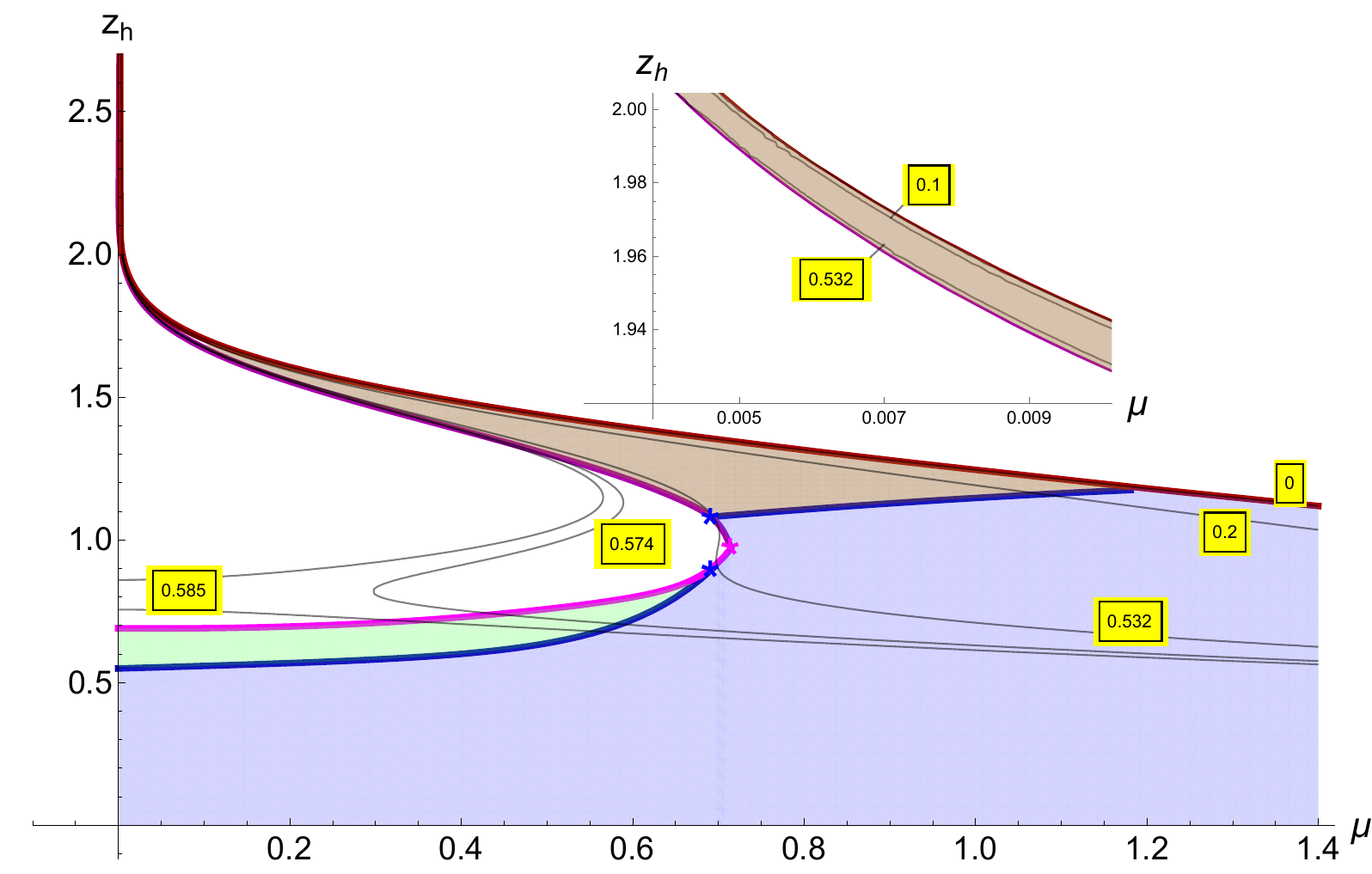}\\
 A\hspace{15em}B
\caption{The light quarks (A) and heavy quarks (B) models. 2D plot in $(\mu,z_h)$-plane with different phases, i.e. QGP, quarkyonic and hadronic corresponding to blue, green and brown regions, respectively.  Solid black lines show the temperature indicated in yellow squares and the intersection of the confinement/deconfinement and the 1st order phase transition lines is denoted by the blue stars. The magenta stars indicate CEPs; $[\mu]=[z_h]^{-1} =$ GeV. 
}
\label{Fig:PhL2D}
\end{figure}

In Fig.\,\ref{Fig:PD-LH-HQ-paint}, the phase structure of the light and heavy quarks models is  presented. The green lines describe the 1st order phase transitions. Below these lines, a hadronic phase is located. The blue lines represent the confinement-deconfinement phase transitions which can be obtained by Wilson loop calculations \cite{Li:2017tdz,Yang:2015aia,Arefeva:2023jjh}. Between the green and blue lines we have a quarkyonic phase, and there is a QGP phase above the blue line. In both cases, the magenta star denotes the critical end point (CEP). Instead of $(\mu,T)$-plane for the phase diagram it is useful to consider $(\mu,z_h)$-plane, i.e. Fig.\,\ref{Fig:PhL2D},  the whole procedure for obtaining which is described in \cite{AHSU}.\\

\newpage

\section{Beta-Function} \label{sec:beta}

Beta-function, $\beta(\alpha)$, encodes the dependence of coupling constant $\alpha$ on the energy scale of the physical system. a holographic $\beta$-function is given by 
\cite{deBoer:1999tgo, Gursoy:2007cb,He:2010ye}
 \be \label{beta-z}
\beta(\alpha)=3\,\alpha \,X \,,
 \ee
where $\alpha(z)$ is defined as \cite{Gursoy:2007cb,Gursoy:2007er,Pirner:2009gr}
\be\label{lambda-phi}
\alpha(z)=e^{\varphi(z)} \,
\ee
and $X$ is a new dynamical variable \cite{Gursoy:2007cb, Gursoy:2007er, Arefeva:2019qen,Arefeva:2018jyu,Arefeva:2020aan} defined as 
\be \label{X-z}
X(z)=\frac{\dot{\varphi} B}{3\dot{B}}\,,
\ee
where $B(z)$ is defined in (\ref{warp-factor}) and $\varphi$ for the light and heavy quarks model are obtained in (\ref{Lphiz}) and \eqref{phiHQ}, respectively.\\

To obtain physically reasonable answers for the $\beta$-function, we need to impose proper boundary conditions on the dilaton field, which also fix the running coupling behavior \cite{AHSU}. In general, the dilaton field with a boundary condition at the holographic coordinate $z=z_0$ is denoted by
\be
\varphi_{z_0}(z)=\varphi_{0}(z)-\varphi_{0}(z_0).\ee
As follows from  Sect.\,\ref{sec:prelim}, the running coupling depends on the boundary conditions,
\bea
&&\alpha_{0}(z)\to\alpha_{z_0}(z)=\alpha_0(z)\,\fG (z_0) \\ &&\mbox{where}\quad\alpha_0(z)=e^{\varphi_0(z)},
\quad\fG (z_0)=e^{-\varphi_{0}(z_0)}
\eea
 and, then $\beta$-function can be written as
 \be\label{beta0}
\beta_0(z)\to   \beta_{z_0}(z)= \beta_0(z)\fG (z_0) \quad\mbox{with}\quad \beta_0(z)=3\alpha_0(z)X(z), 
 \ee
here the function $X(z)$ does not depend on the choice of boundary conditions.
\\

While $\varphi_{0}(z)$ is independent of the thermodynamic quantities such as $T$ and $\mu$  within the model both for the light and heavy quarks, it is possible to cover this dependence for the running coupling expressing $z_0$ in terms of $z_h$, i.e. $z_0=\fz(z_h)$, and get
\be\label{alpha_gen}
\alpha_{\fz}(z;T,\mu)=\alpha_0(z)\,\fG (T,\mu),\quad\mbox{where}\quad\fG (T,\mu)= e^{-\varphi_{0}(\fz(z_h))}
\ee
One such choice is
$\fz(z_h)=z_h$, and the other is given by exponential functions, see \eqref{phi-fz-LQ} and \eqref{bceHQ},
which are different for the light and heavy quarks, respectively.\\

\subsection{Beta-function for the light quarks model as a function of the thermodynamic parameters  }\label{BF-LQ}

\subsubsection {$\beta$-function with the boundary condition $z_0=z_h$ }

 Respecting the first boundary condition (\ref{bch}) which provides thermodynamics properties, the $\beta$-function can be written as
\be\label{beta_LQ_zh}
\beta_{z_h}(z;T,\mu)=\beta_0(z)\,\fG (T,\mu)\quad\mbox{where}\quad\fG (T,\mu)= e^{-\varphi_{0}(z_h)},
\ee
and $\beta_0$ is given by (\ref{beta0}). In general, to produce calculations for the $\beta$-function we need to respect the physical domains of the theory, that is presented in Fig.\,\ref{Fig:PhL2D}A for the light quarks. \\

\begin{figure}[h!]
  \centering
\includegraphics[scale=0.32]{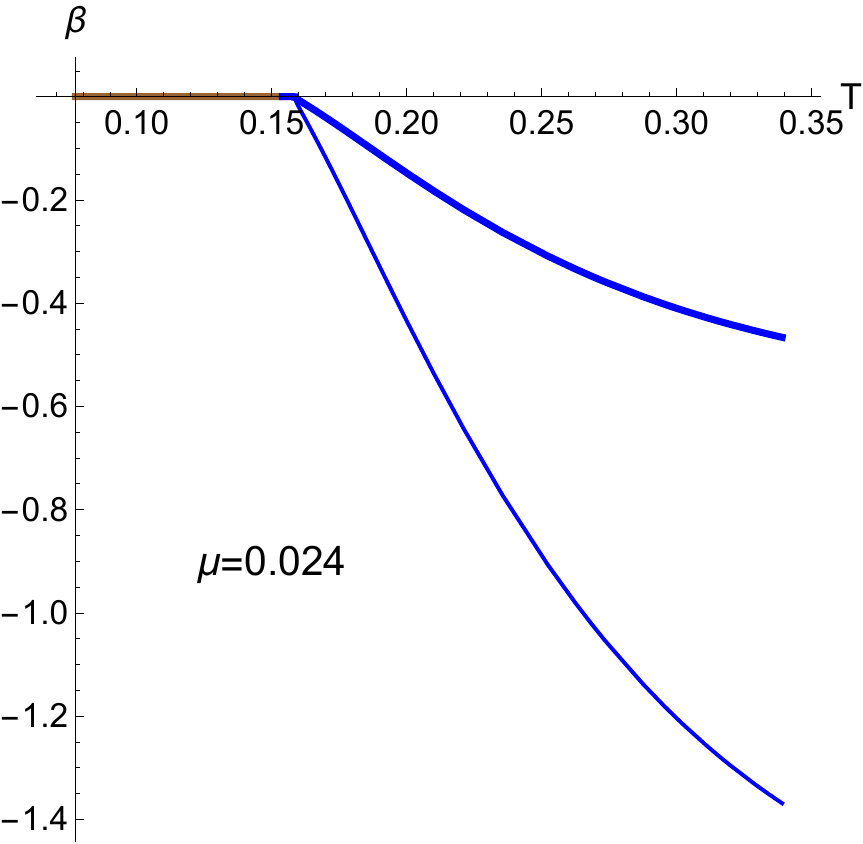} \qquad \qquad \qquad
\includegraphics[scale=0.32]{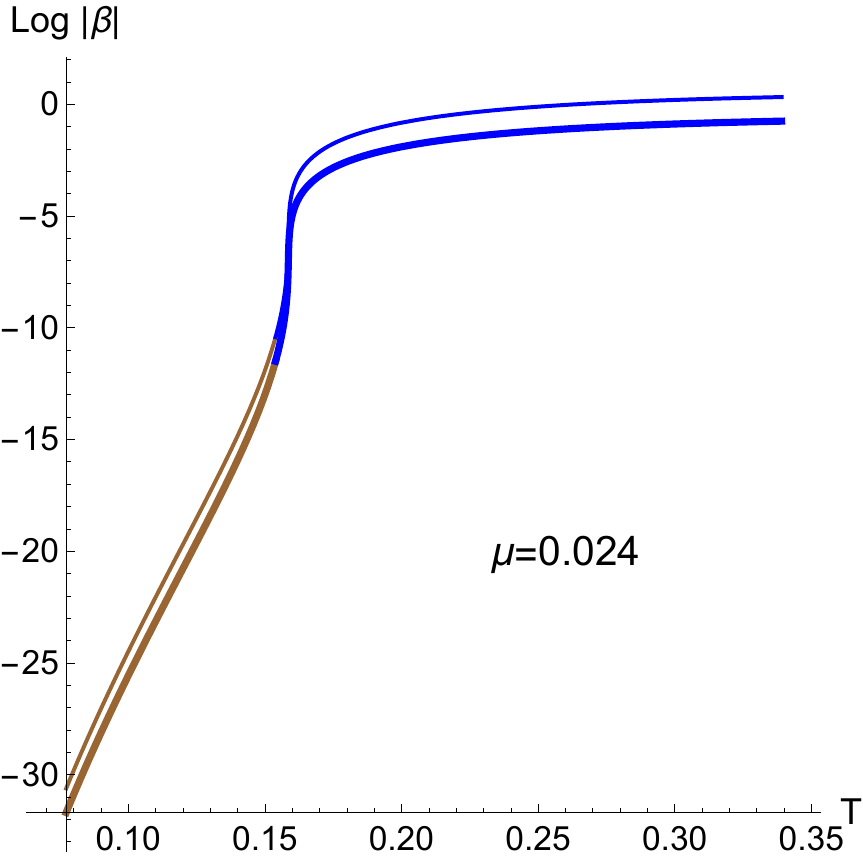}\\
 A\hspace{200pt}B \\\,\\
\includegraphics[scale=0.32]{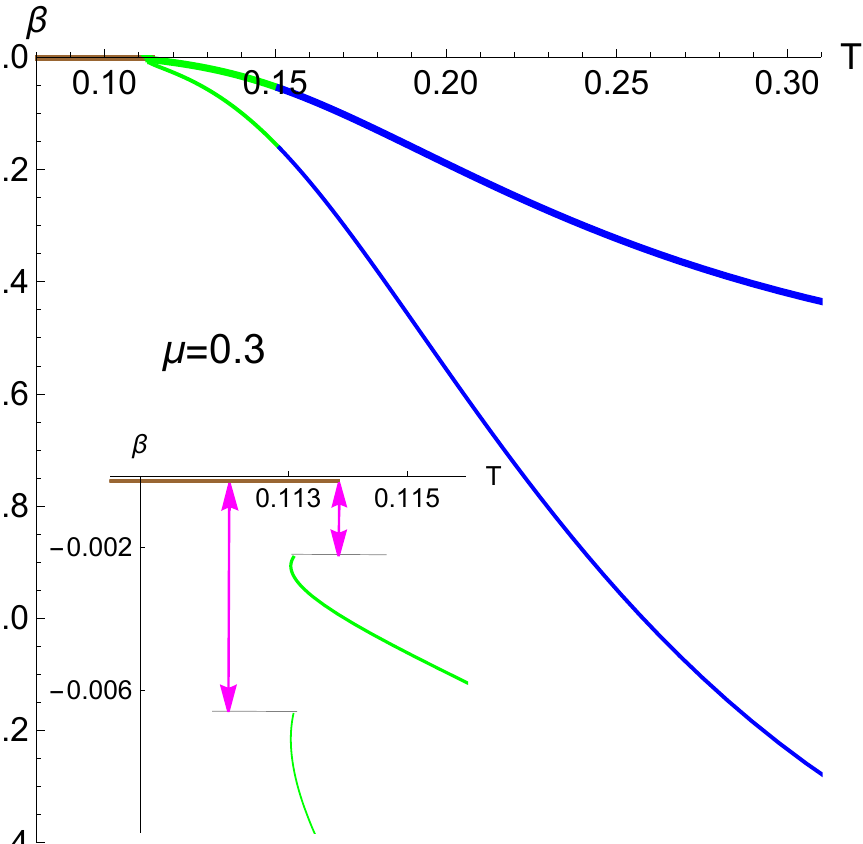} \qquad \qquad \qquad
\includegraphics[scale=0.34]{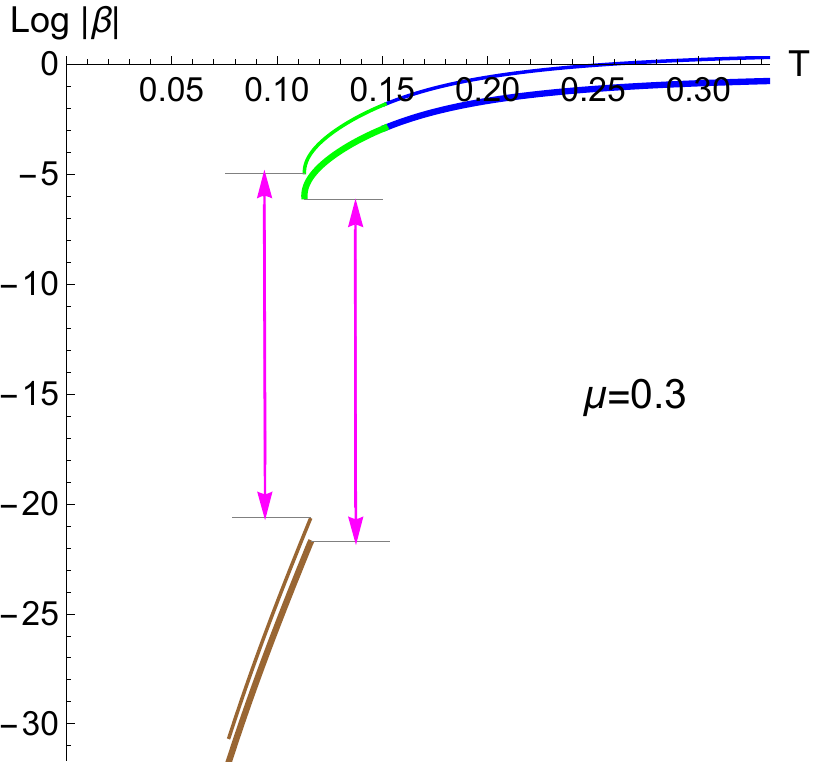}\\
 C\hspace{200pt}D \\\,\\
\includegraphics[scale=0.32]{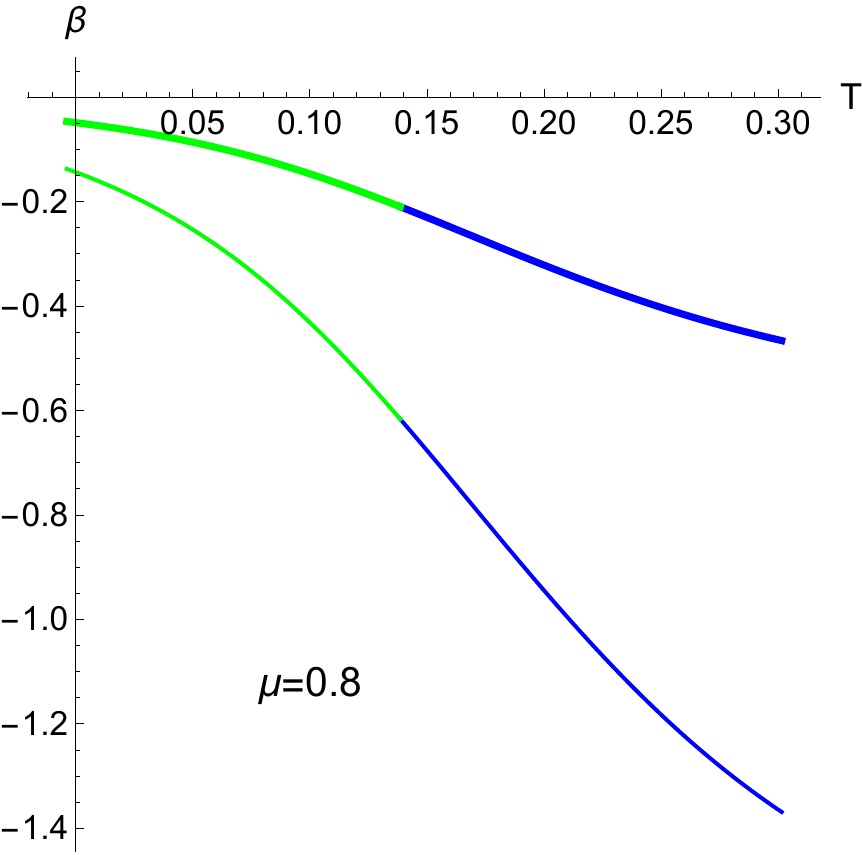} \qquad \qquad  \qquad
\includegraphics[scale=0.31]{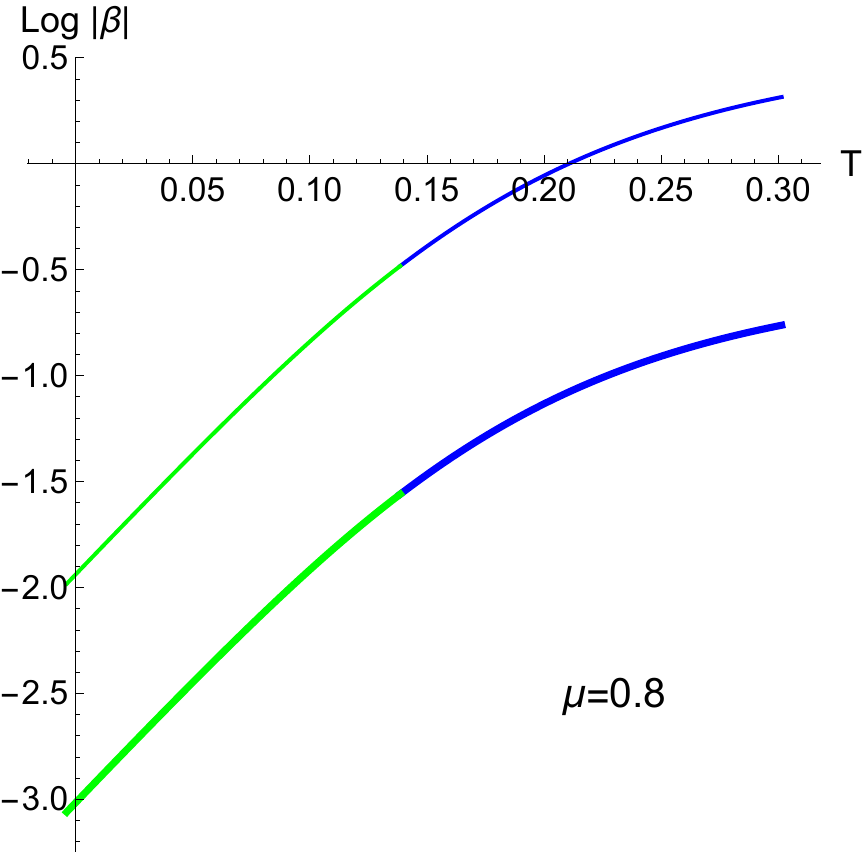}\\
 E\hspace{200pt}F \\
 \,
\caption{Beta-function $\beta=\beta_{z_h}(z;\mu,T)$ and  $ \log|\beta|=\log \, |\beta_{z_h}|(z;\mu,T)$ for the light quarks at fixed $\mu$ at different energy scales  $z=1$ (thin lines) and $z=0.3$ (thick lines).  Hadronic, QGP and quarkyonic phases are denoted by brown, blue and green lines, respectively. Magenta arrows in (C, D) show the jumps at the 1st order phase transition. $[\mu]=[T]=[z]^{-1} =$ GeV. 
}
\label{Fig:LQ-beta2-2D}
\end{figure}

The temperature dependence of the $\beta$-function, $\beta_{z_h}(z;\mu,T)$ and  $\log \, |\beta_{z_h}(z;\mu,T)|$ for the light quarks model at fixed $\mu=0.024$ GeV (A,B), $\mu=0.3$ GeV (C,D) $\mu=0.8$ GeV (E,F) and different energy scales  $z=1$ GeV${}^{-1}$ (thin lines) and $z=0.3$ GeV${}^{-1}$ (thick lines) is depicted in Fig.\,\ref{Fig:LQ-beta2-2D}.  Hadronic, QGP and quarkyonic phases are denoted by brown, blue and green lines, respectively. Although the first boundary condition $z_0=z_h$ is not physical, the $\beta$-function senses a  jump at the 1st order phase transition between the 
 hadronic and quarkyonic phases, and the  magenta arrows represent this event in (C,D). In $\mu=0.024$ GeV (A) and $\mu=0.8$ GeV (E), there is a phase transition without any jump between hadronic to QGP and between quarkyonic to QGP, respectively. The $\beta$-function decreases with increasing $T$ at a fixed $\mu$. At low values of the temperature, i.e. at the hadronic phase, the $\beta$-function becomes very small but it is not exactly zero.


\subsubsection {$\beta$-function with the  boundary condition $z_0=\fz_{LQ}(z_h)$ } \label{LQ-nbc10}

Similar to (\ref{beta_LQ_zh}), the $\beta$-function with the second boundary condition (\ref{phi-fz-LQ}) for the light quarks can be written as
\be\label{beta_LQ_fz}
\beta_{\fz_{_{LQ}}}(z;T,\mu)=\beta_0(z)\,\fG (\fz_{_{LQ}}(T,\mu))\quad\mbox{where}\quad\fG (\fz_{_{LQ}}(T,\mu))= e^{-\varphi_{0}(\fz_{_{LQ}}(z_h(T,\mu)))},
\ee
and $\beta_0$ is given by (\ref{beta0}). Note that this boundary condition is physical one. \\

\begin{figure}[b!]
  \centering
\includegraphics[scale=0.35]{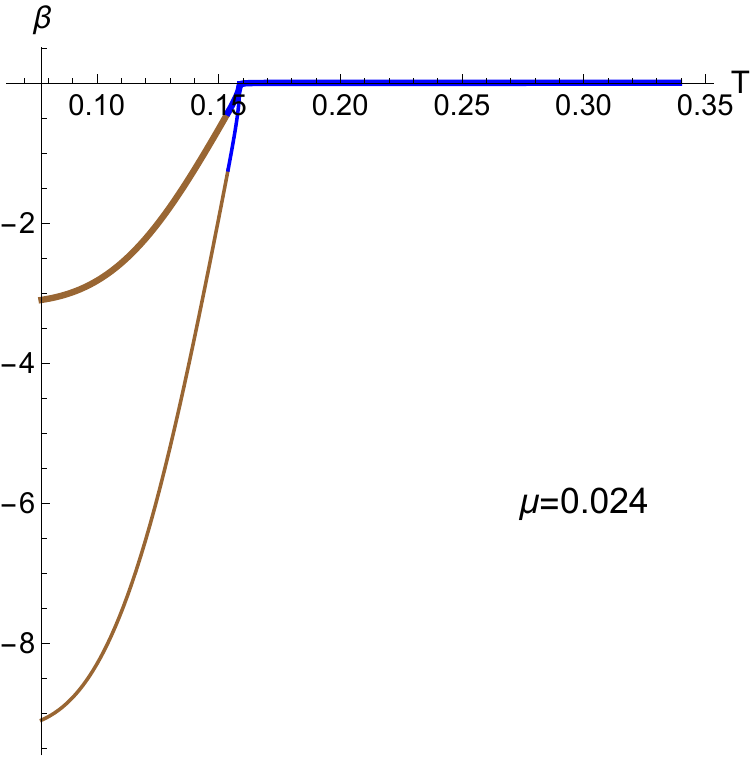} \qquad \qquad \qquad
\includegraphics[scale=0.35]{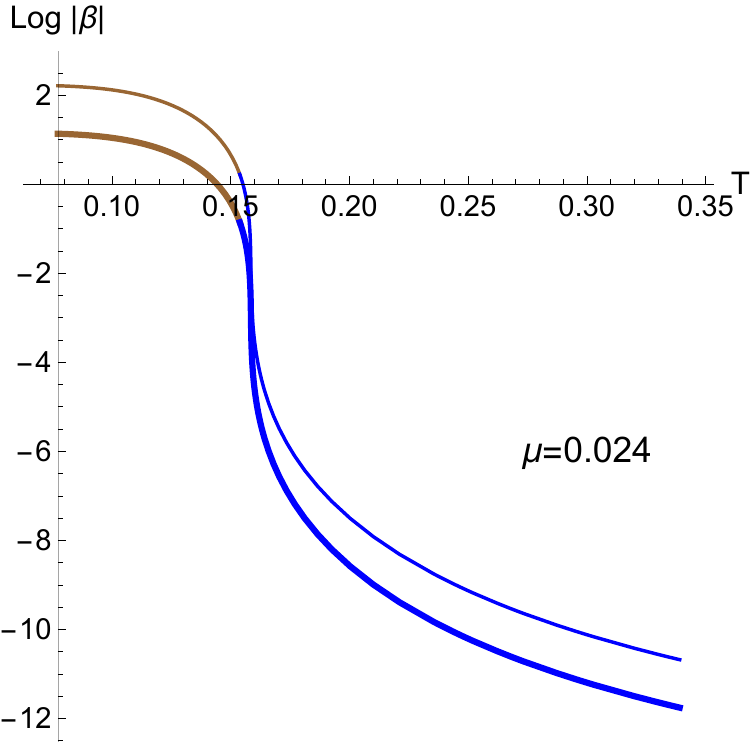}\\
 A\hspace{200pt}B \\
\includegraphics[scale=0.35]{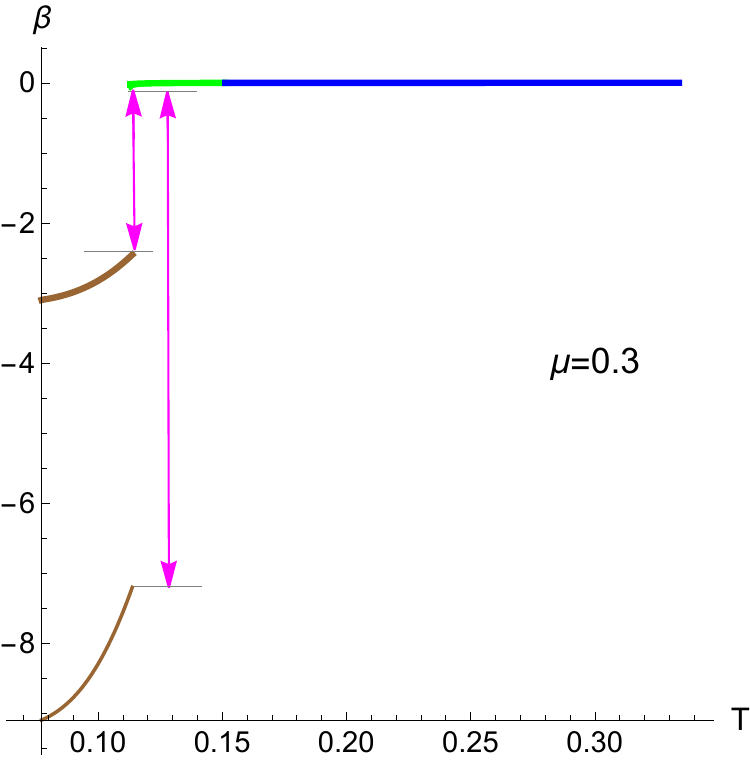} \qquad \qquad \qquad
\includegraphics[scale=0.35]{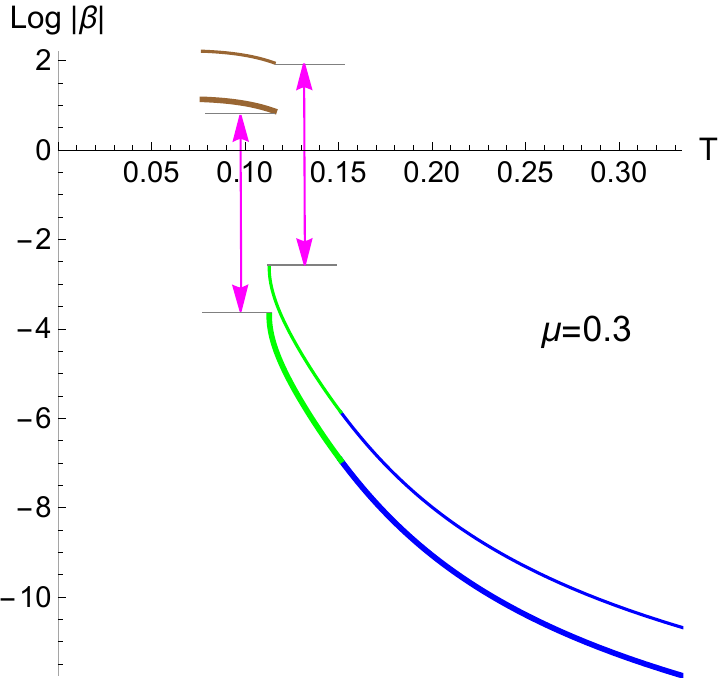}\\
 C\hspace{200pt}D \\
\includegraphics[scale=0.36]{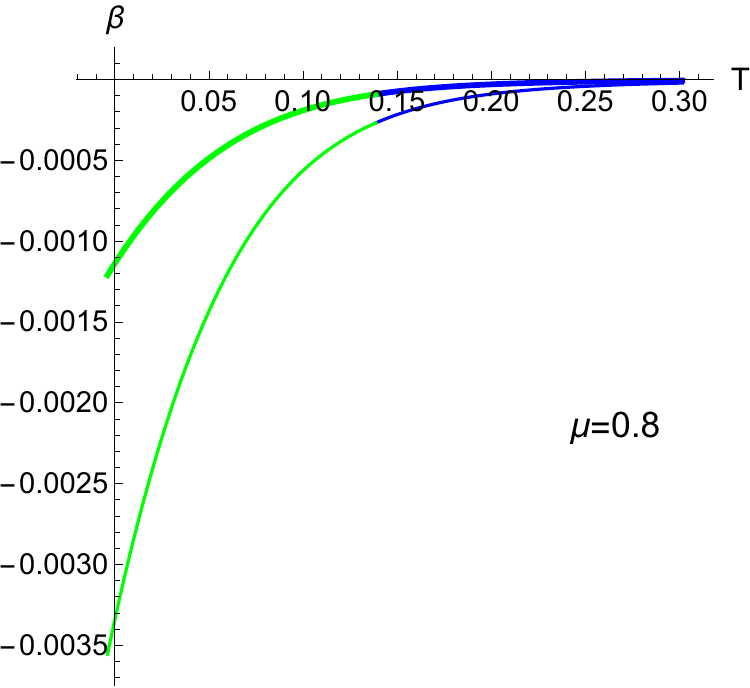} \qquad \qquad  \qquad
\includegraphics[scale=0.35]{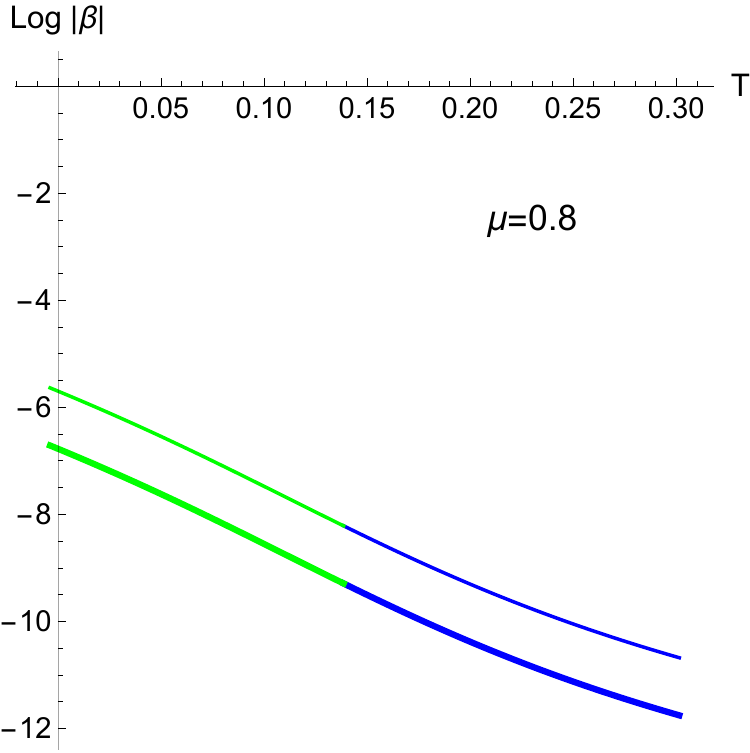}\\
 E\hspace{200pt}F
\caption{Beta-function $\beta=\beta_{\fz_{_{LQ}}}(z;\mu,T)$ and  $ \log|\beta|=\log \,|\beta_{\fz_{_{LQ}}}(z;\mu,T)|$ for the light quarks at fixed $\mu$ at different energy scales  $z=1$ (thin lines) and $z=0.3$ (thick lines).  Hadronic, QGP and quarkyonic phases are denoted by brown, blue and green lines, respectively. Magenta arrows in (C, D) show the jumps at the 1st order phase transition. $[\mu]=[T]=[z]^{-1} =$ GeV. 
}
\label{Fig:LQ-beta2-2D-z}
\end{figure}

The temperature dependence of $\beta_{\fz_{_{LQ}}}(z;\mu,T)$   and 
$ \log \,|\beta_{\fz_{_{LQ}}}(z;\mu,T)|$
for the light quarks model at fixed $\mu=0.024$ GeV (A,B), $\mu=0.3$ GeV (C,D) $\mu=0.8$ GeV (E,F) and different energy scales  $z=1$ GeV${}^{-1}$ (thin lines) and $z=0.3$ GeV${}^{-1}$ (thick lines) are depicted in Fig.\,\ref{Fig:LQ-beta2-2D-z}.  Hadronic, QGP and quarkyonic phases are denoted by brown, blue and green lines, respectively. Magenta arrows in C and D show that the $\beta$-function feels the jumps at the 1st order phase transition between the  hadronic and quarkyonic phases. In $\mu=0.024$ GeV (A) and $\mu=0.8$ GeV (E), there is a phase transition without any jump between hadronic to QGP and between quarkyonic to QGP, respectively. The $\beta$-function increases with increasing $T$ for a fixed $\mu$. This behavior is obtained utilizing the physical boundary condition (\ref{phi-fz-LQ}). Note that due to the dependence of $\alpha$ on the boundary condition, the behavior of the $\beta$-function in terms of the 
 thermodynamic quantities will change as well in comparison with Fig.\,\ref{Fig:LQ-beta2-2D}.\\

 \newpage

\subsection{Beta-Function for the heavy quarks model as a function of thermodynamic parameters}\label{BF-HQ}

\subsubsection {$\beta$-function with the 
 boundary condition $z_0=z_h$ }

\begin{figure}[b!]
  \centering
\includegraphics[scale=0.30]{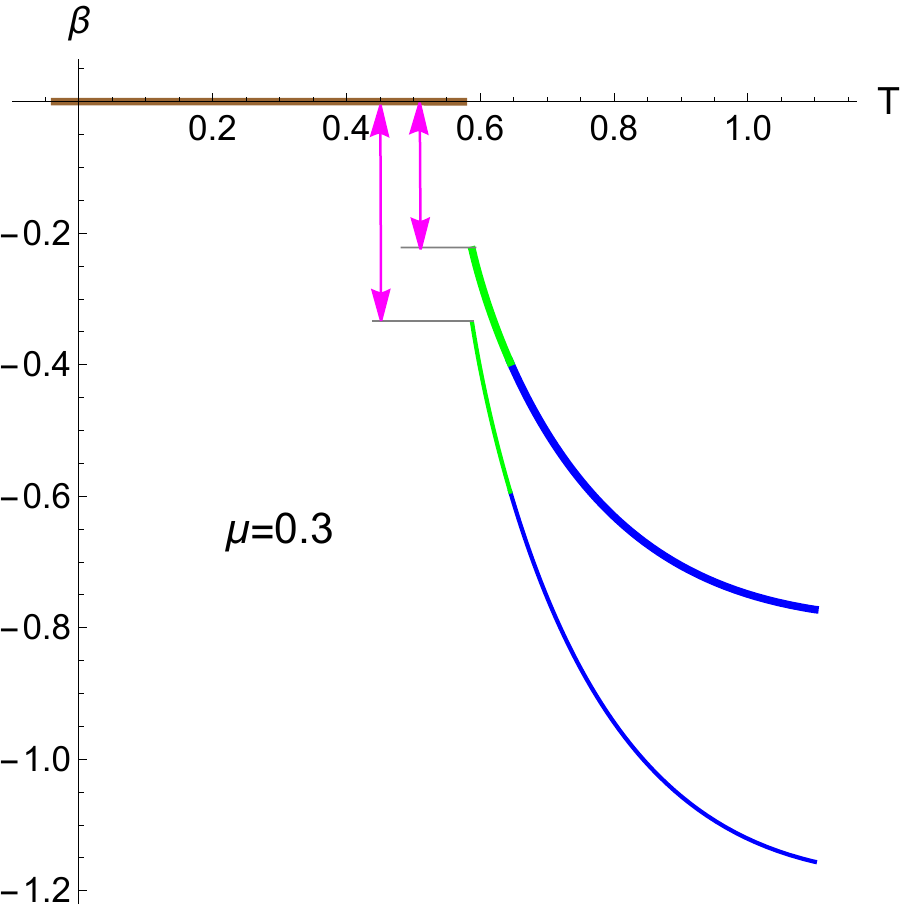} \qquad \qquad \qquad
\includegraphics[scale=0.31]{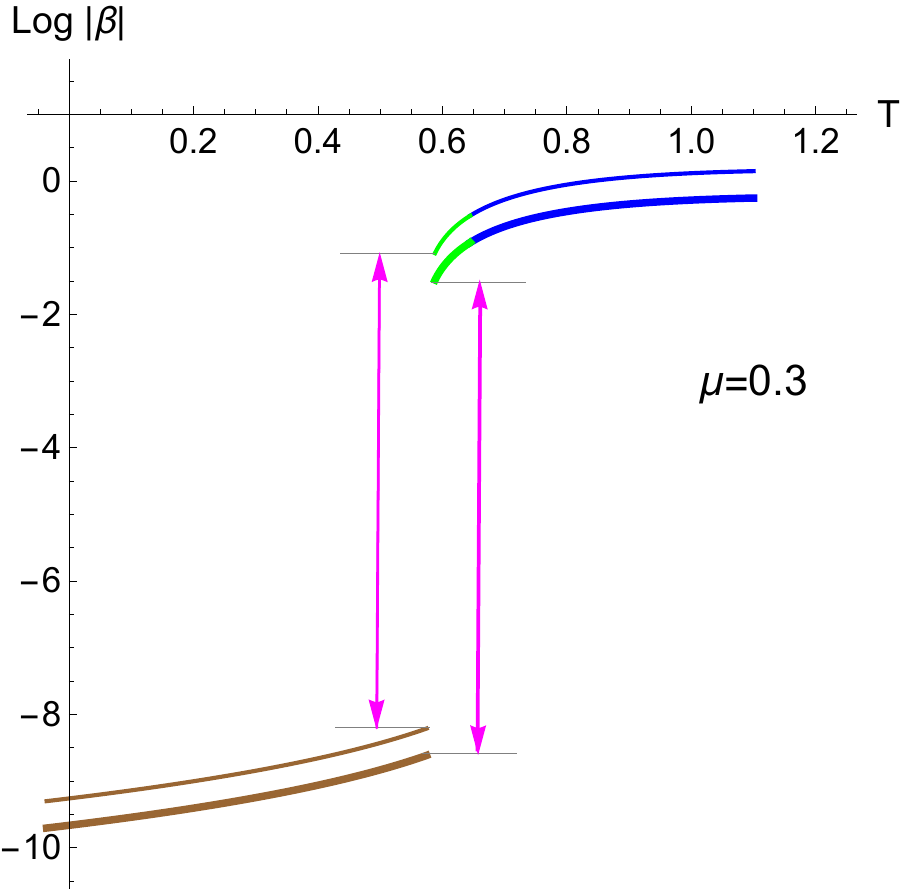}\\
 A\hspace{200pt}B \\\,\\
\includegraphics[scale=0.31]{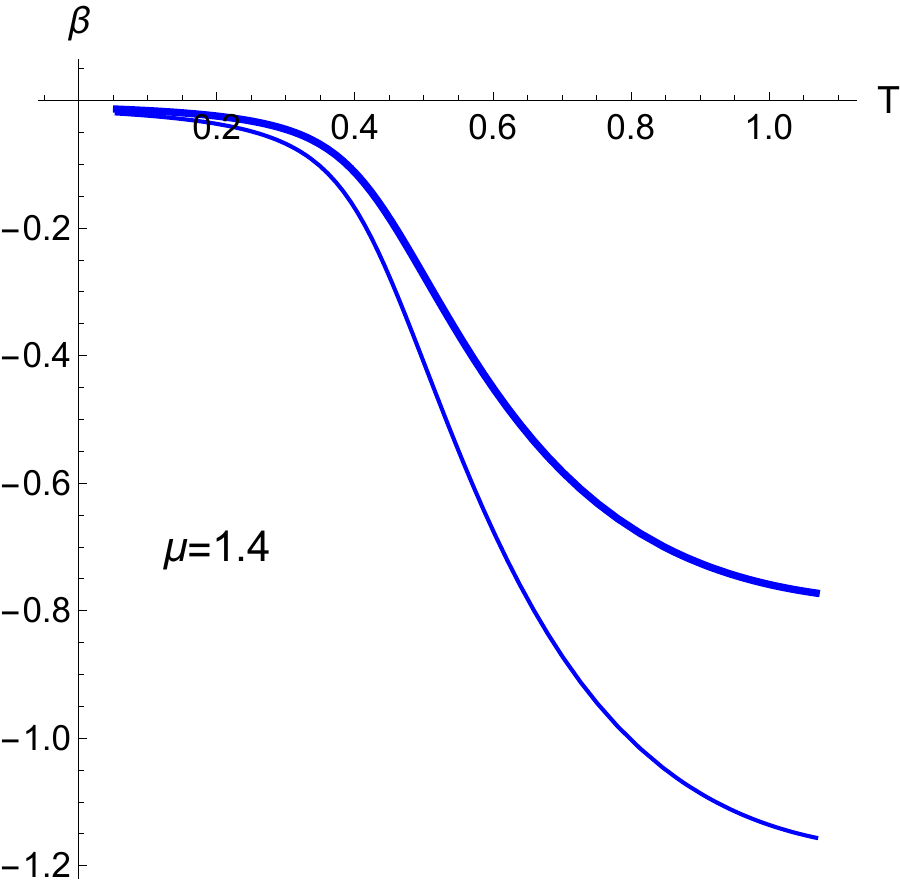} \qquad \qquad \qquad
\includegraphics[scale=0.31]{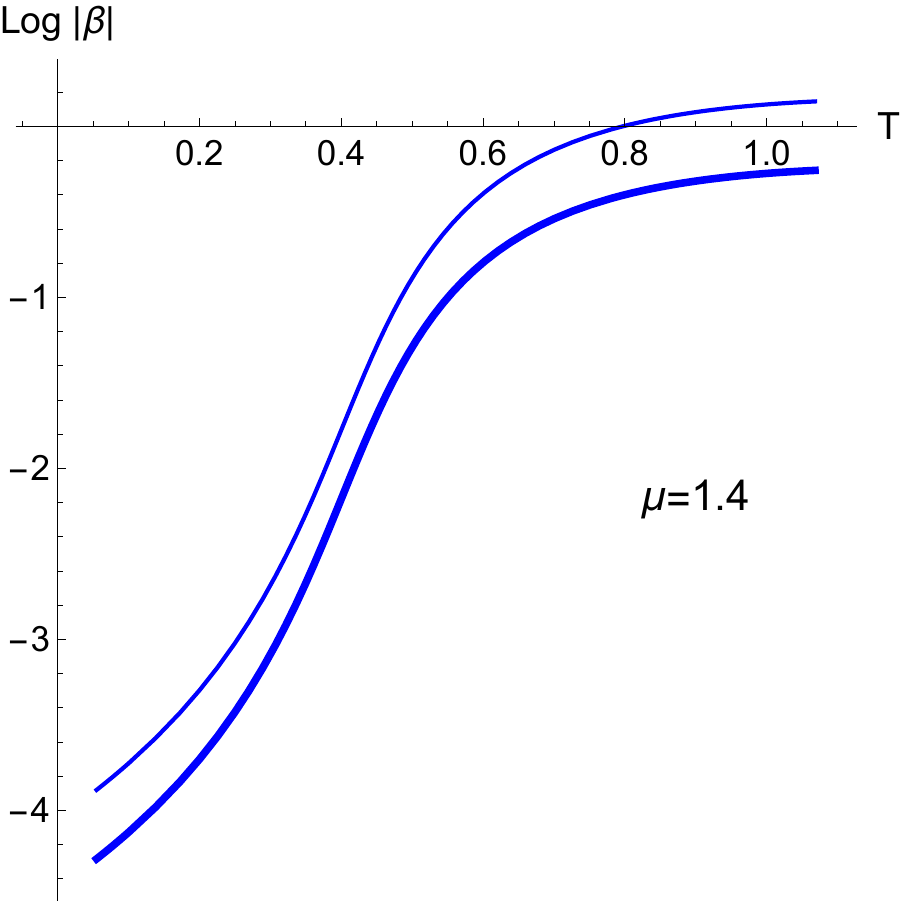}\\
 C\hspace{200pt}D \\\,\\
 \,
\caption{ Beta-function $\beta=\beta_{z_h}(z;\mu,T)$ and  $ \log|\beta|=\log \,|\beta_{z_h}(z;\mu,T)|$ for the heavy quarks at fixed $\mu$ at different energy scales  $z=0.3$ (thin lines) and $z=0.2$ (thick lines).  Hadronic, QGP and quarkyonic phases are denoted by brown, blue and green lines, respectively. Magenta arrows in (A,B) show the jumps at the 1st order phase transition. $[\mu]=[T]=[z]^{-1} =$ GeV. 
}
\label{Fig:HQ-beta-2D}
\end{figure}

Applying the first boundary condition (\ref{bch}) for the heavy quarks model, the $\beta$-function takes the form
\be\label{beta_HQ_zh}
\beta_{z_h}(z;T,\mu)=\beta_0(z)\,\fG (T,\mu)\quad\mbox{where}\quad\fG (T,\mu)= e^{-\varphi_{0}(z_h)},
\ee
and $\beta_0$ is given by (\ref{beta0}). For the heavy quarks calculations, we respect the physical domains of the theory presented in Fig.\,\ref{Fig:PhL2D}B. \\

The temperature dependence of the $\beta$-function, $\beta_{z_h}(z;\mu,T)$ and  $ \log \,|\beta_{z_h}(z;\mu,T)|$ for the heavy quarks model at fixed $\mu=0.3$ GeV (A,B), $\mu=1.4$ GeV (C,D)  at different energy scales  $z=0.3$ GeV${}^{-1}$ (thin lines) and $z=0.2$ GeV${}^{-1}$ (thick lines) are depicted in Fig.\,\ref{Fig:HQ-beta-2D}.  Hadronic, QGP and quarkyonic phases are denoted by brown, blue and green lines, respectively. Magenta arrows in (A,B) show that the jumps at the 1st order phase transition between the  hadronic and quarkyonic phases.
In $\mu=1.4$ GeV (C), there is no phase transition because the QGP phase is dominated. Similar to the case of the light quarks model with the unphysical boundary condition, i.e. Fig.\,\ref{Fig:LQ-beta2-2D}, the $\beta$-function in Fig.\,\ref{Fig:HQ-beta-2D} is negative and decreases with increasing $T$ at a fixed $\mu$.

\subsubsection {$\beta$-function with the  boundary condition $z_0=\fz_{HQ}(z_h)$ } 
\label{HQ-nbc15}

 The $\beta$-function with the second (physical) boundary condition for the heavy quarks (\ref{bceHQ}) is given by
\be\label{beta_HQ_fz}
\beta_{\fz_{_{HQ}}}(z;T,\mu)=\beta_0(z)\,\fG (\fz_{_{HQ}}(T,\mu))\quad\mbox{where}\quad\fG (\fz_{_{HQ}}(T,\mu))= e^{-\varphi_{0}(\fz_{_{HQ}}(z_h(T,\mu)))},
\ee
and $\beta_0$ is defined in (\ref{beta0}). \\
\begin{figure}[h!]
  \centering
\includegraphics[scale=0.40]{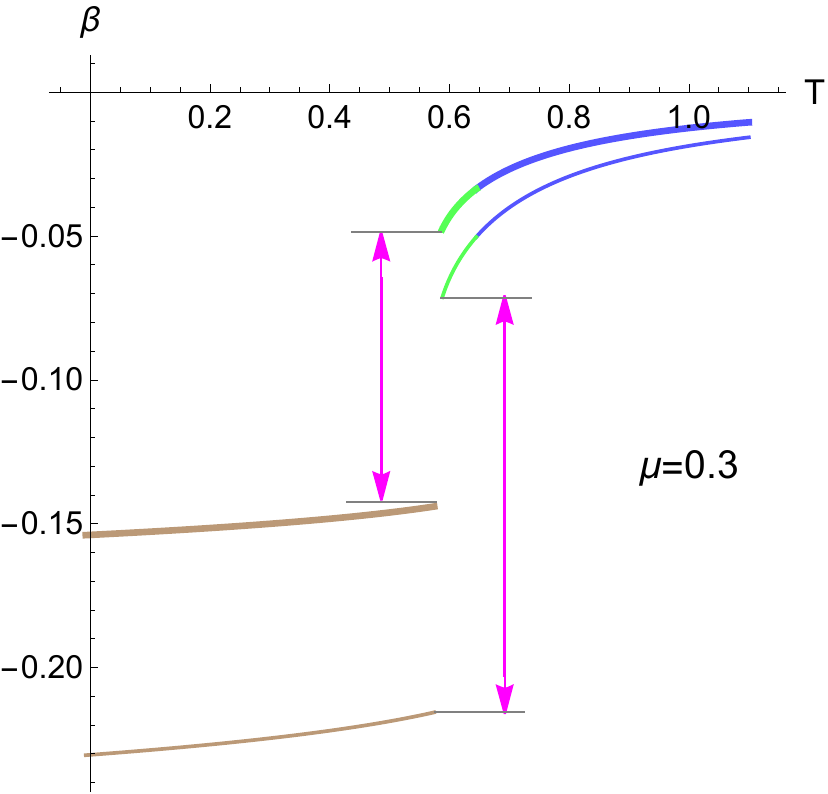} \qquad \qquad \qquad
\includegraphics[scale=0.40]{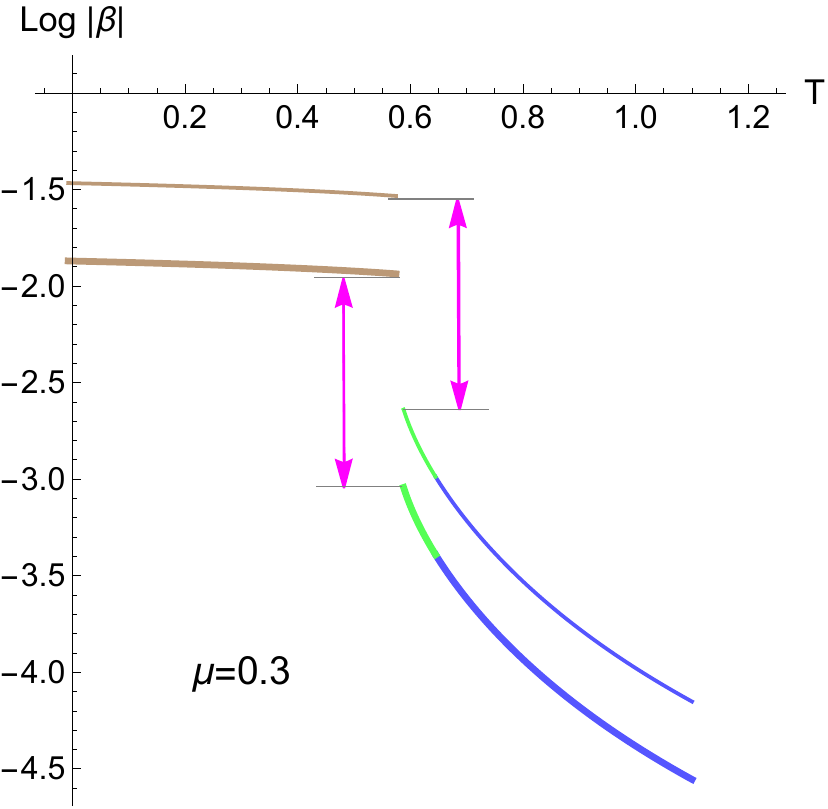}\\
 A\hspace{200pt}B \\\,\\
\includegraphics[scale=0.40]{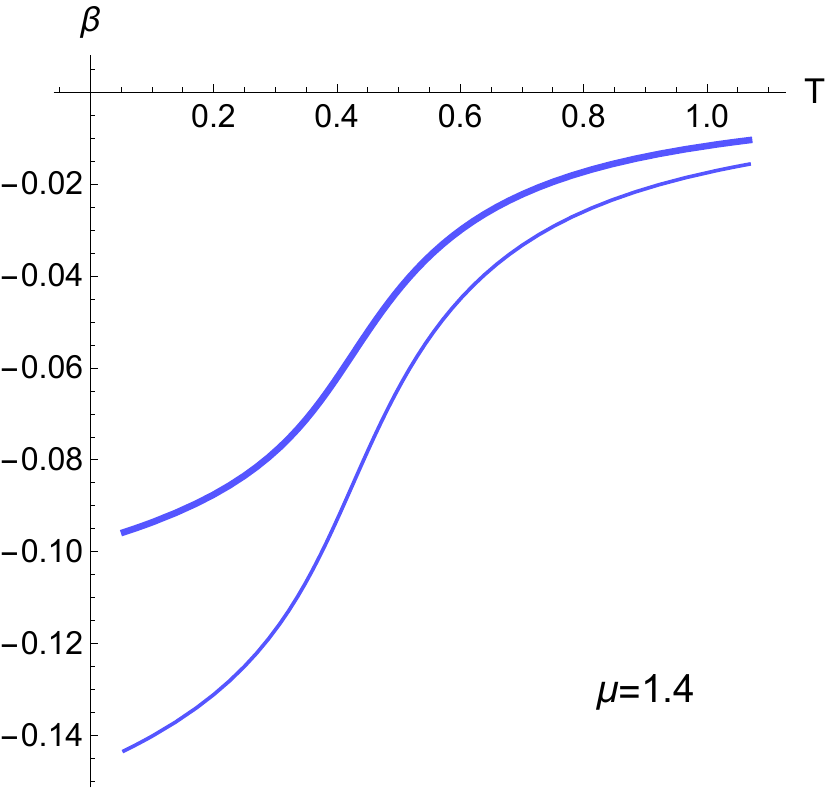} \qquad \qquad \qquad
\includegraphics[scale=0.40]{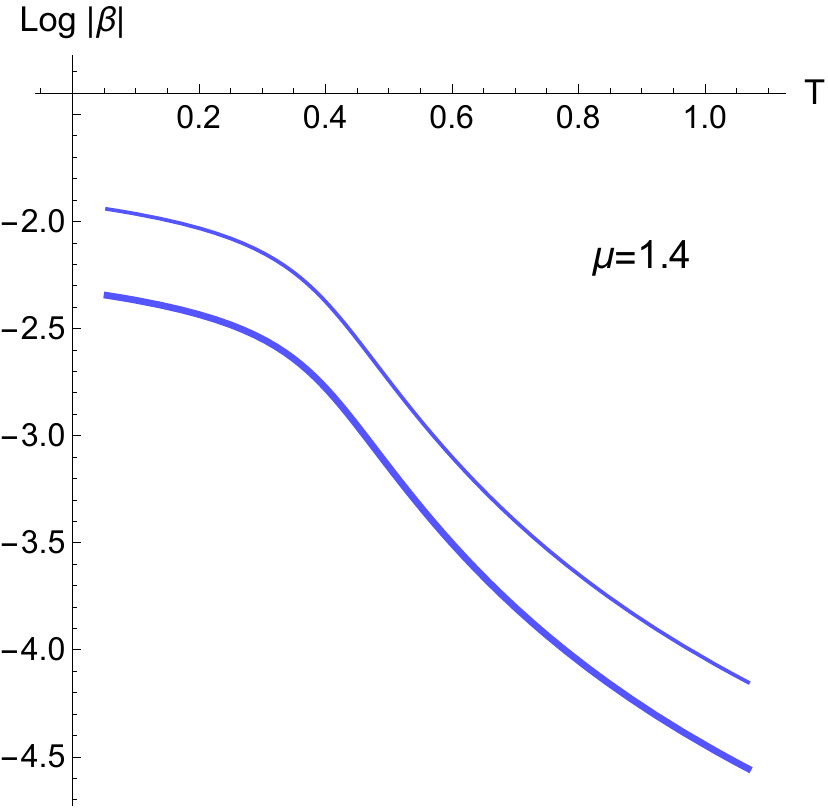}\\
 C\hspace{200pt}D 
\caption{Beta-function $\beta=\beta_{\fz_{_{HQ}}}(z;\mu,T)$ and  $ \log|\beta|=\log \,|\beta_{\fz_{_{HQ}}}(z;\mu,T)|$ for the heavy quarks at fixed $\mu$ at different energy scales  $z=0.3$ (thin lines) and $z=0.2$ (thick lines).  Hadronic, QGP and quarkyonic phases are denoted by brown, blue and green lines, respectively. Magenta arrows in (A,B) show the jumps at the 1st order phase transition. $[\mu]=[T]=[z]^{-1} =$ GeV. 
}
\label{Fig:LQ-beta-2Dgbc}
\end{figure}

The temperature dependence of the $\beta$-function, $\beta_{\fz_{_{HQ}}}(z;\mu,T)$ and  $ \log \,|\beta_{\fz_{_{HQ}}}(z;\mu,T)|$ for the heavy quarks model at fixed $\mu=0.3$ GeV (A,B), $\mu=1.4$ GeV (C,D)  at different energy scales  $z=0.3$ GeV${}^{-1}$ (thin lines) and $z=0.2$ GeV${}^{-1}$ (thick lines) are depicted in Fig.\,\ref{Fig:LQ-beta-2Dgbc}.  Hadronic, QGP and quarkyonic phases are denoted by brown, blue and green lines, respectively. Magenta arrows in (A,B) show that the jumps at the 1st order phase transition between the  hadronic and quarkyonic phases.
In $\mu=1.4$ GeV (C), there is no phase transition because the QGP phase is dominated.  Considering the physical boundary condition, the $\beta$-function is negative and increases with increasing $T$ at a fixed $\mu$. The qualitative behavior of the $\beta$-function for the light and heavy quarks models are the same for each type of the boundary conditions. \\


$$\,$$
\newpage
$$\,$$
\newpage

\section{RG flow} \label{sec:RG}

\subsection{RG flow equation for the  zero temperature and zero chemical potential }

Let us consider a  zero temperature and zero chemical potential case. One  defines a new dynamical variable \cite{ Gursoy:2007cb, Gursoy:2007er, Arefeva:2019qen,Arefeva:2018jyu,Arefeva:2020aan}
\be \label{X-z}
X(z)=\frac{\dot{\varphi} B}{3\dot{B}},
\ee
and consider $\cX$ such that
\be
X(z)=\cX(\varphi(z)).\ee
Following from the EOMs (\ref{phi2primeT0mu0})-(\ref{VT0mu0}), the function $\cX(\varphi)$
 satisfies the equation
 \bea\label{cXphi}
\frac{d\cX}{d\varphi}&=&-\frac{4}{3} \left(1-\frac{3 }{8}\cX^2\right) \left(1+\frac{1}{\cX}\frac{\partial_{\varphi}\cV(\varphi)}{ \cV(\varphi)}\right).
\eea 
One calls this equation the RG flow equation \cite{Gursoy:2007cb, Gursoy:2007er}, since 
$\cX$ is related with the
 $\beta$-function defined as (\ref{beta-z}).\\

The dilaton field $\varphi$ with the zero boundary condition at the zero holographic coordinate is
\be
\varphi_0(z)\Big|_{z=0}=0.\ee
In addition, one can consider the boundary condition 
\be
\varphi_0(z)\Big|_{z=0}=-10,\ee
to cover $\alpha<1$.

\subsubsection{RG flow for the light quarks model}

 \begin{figure} [h!]
 \centering \includegraphics[scale=0.36]{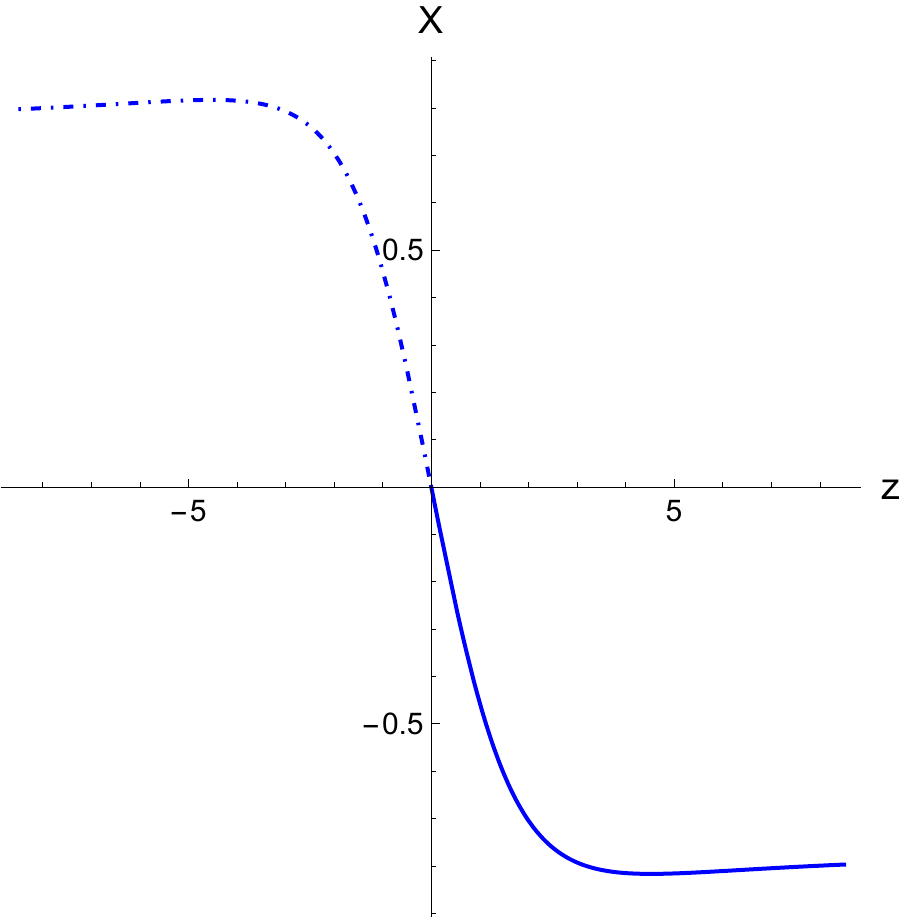} \quad 
 \includegraphics[scale=0.35]{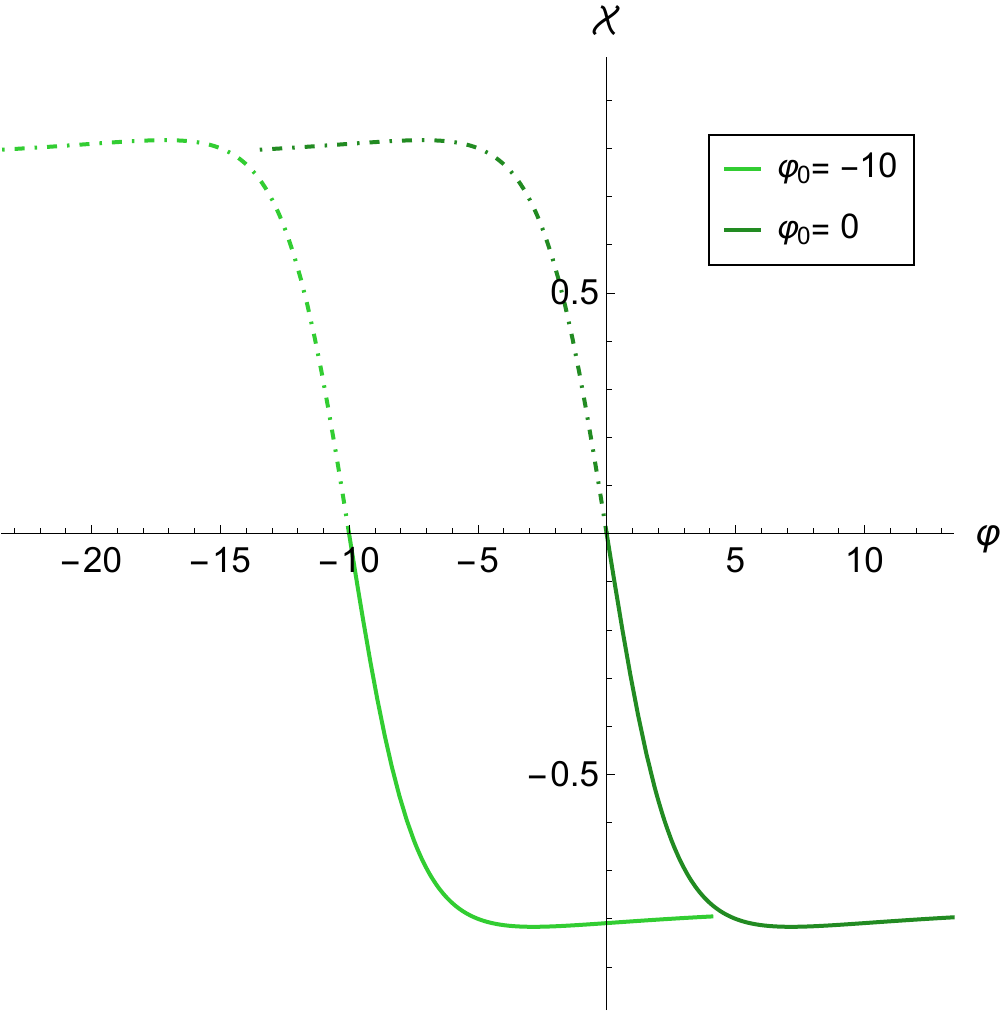}\\
A\hspace{200.pt}B
\caption{The behavior of $X=X(z)$ for the light quarks (A), and $\cX=\cX(\varphi)$ with different boundary conditions, i.e.  $\varphi_0=-10$ (lime line) and $\varphi_0=0$ (green line) for the light quarks (B). The dotdashed line represents an unphysical region $z<0$ ; $[z]^{-1} =$ GeV. 
}
    \label{fig:LQXzisoT=0mu=0}
\end{figure}
The plot of $X(z)$ given by the equation \eqref{X-z} is presented in Fig.\,\ref{fig:LQXzisoT=0mu=0}A, and $\cX=\cX(\varphi)$ is presented in Fig.\,\ref{fig:LQXzisoT=0mu=0}B with different boundary conditions, i.e.  $\varphi_0=-10$ (lime line) and $\varphi_0=0$ (green line). 
The graphs in Fig.\,\ref{fig:LQXzisoT=0mu=0}B are recovered from equations (\ref{X-z}), (\ref{Lphiz}) and correspond to exact solutions of the light quarks model. We see that in Fig.\,\ref{fig:LQXzisoT=0mu=0}A,  $X(z) \to 0$ for $z \to 0 $ and in Fig.\,\ref{fig:LQXzisoT=0mu=0}B , ${\cal X}(\varphi) \to 0$ for $\varphi \to \varphi_0 $. This happens because   $\dot{\varphi} \to 0$ and $ B/\dot{B} \to 0$ in \eqref{X-z} when $z \to 0$.   \\

\begin{figure}[h!]  
\centering
\hspace{5em} \includegraphics[scale=0.44]{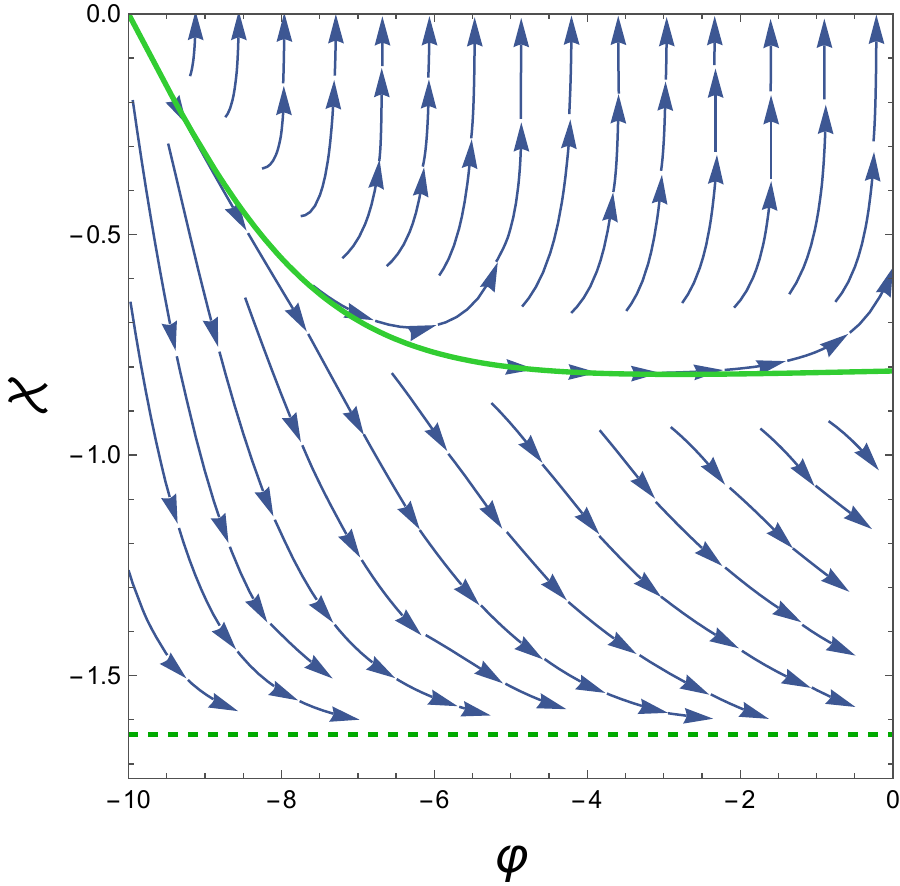}
 \quad
\includegraphics[scale=0.45]{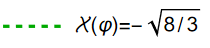}
\caption{RG flows for the light quarks in isotropic background with $T=0$ and $\mu=0$ corresponding to $\cV(\varphi)$ with  
boundary condition  $\varphi_0=-10$. 
 The green dashed line represents $\cX=- \sqrt{8/3}$ fixed (atractor) line.
 The green solid line represents $\cX(\varphi)$ shown in Fig.\,\ref{fig:LQXzisoT=0mu=0}B.} 
 \label{Fig:RGLQisomu0T0}
 \end{figure}
 
 RG flow  for the case $T=0$ and $\mu=0$  with the boundary condition $\varphi_0=-10$ corresponding to an approximation of the dilaton potential $\cV(\varphi)$, namely (\ref{VL}), is shown in Fig.\,\ref{Fig:RGLQisomu0T0}.
 The green dashed line represents $\cX=- \sqrt{8/3}$ fixed (attractor) line.  Here we considered just a physical region $z>0$. The Fig.\,\ref{Fig:RGLQisomu0T0} shows that our solution (the solid lime line) is in correspondence with repulsive behavior of RG flows. Therefore, our solution is unstable that is associated with the negative potential.
In Fig.\,\ref{Fig:RGLQisomu0T0} we see the comparison between our solution and 
other solutions of \eqref{cXphi}. Also ${\cal X}(\varphi) \to 0$ when $\varphi \to \varphi_0 $ and ${\cal X}(\varphi) \to const$  when $\varphi \to 0 $.  In Fig.\,\ref{Fig:RGLQisomu0T0} the approximation for the potential is checked  only for $-23.5 \leq \varphi \leq 3.5$, see Fig.\,\ref{Fig:VPhiHisomu=0T=0}A. So we can trust  the   stream plot  only for  $|\varphi+10| \leq 13$.\\

\subsubsection{RG flow for the heavy quarks model}

\begin{figure}[h!]
 \centering \includegraphics[scale=0.34]{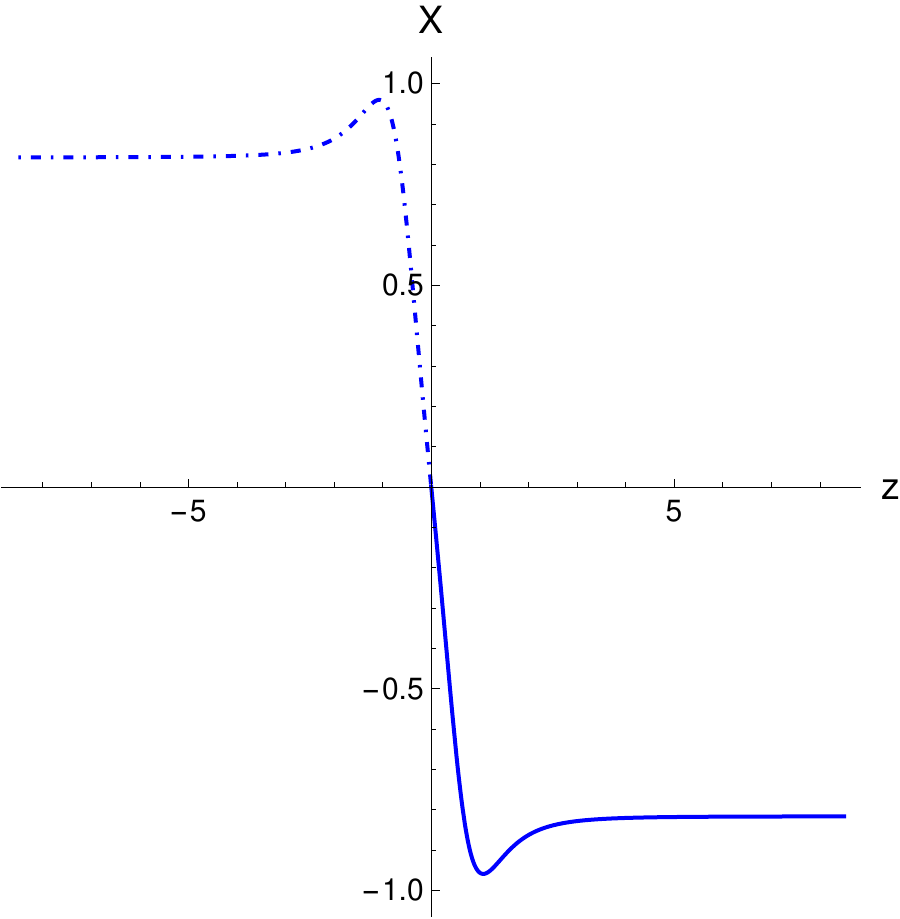} \quad \includegraphics[scale=0.32]{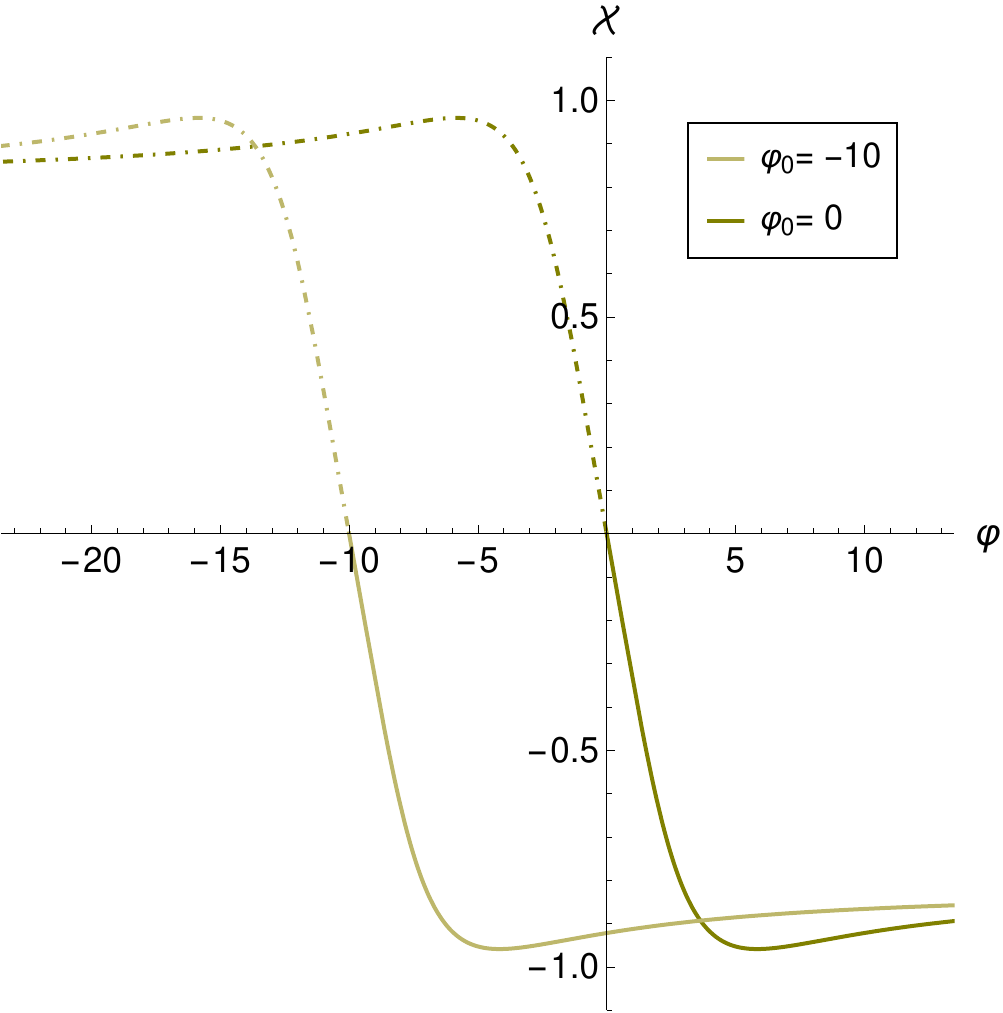}\\
 A\hspace{200.pt}B
\caption{The behavior of (A) $X=X(z)$, and  (B) $\cX=\cX(\varphi)$ with different boundary condition $\varphi_0=-10$ (khaki line) and $\varphi_0=0$ (olive line) for the heavy quarks. The dotdashed line represents an unphysical region $z<0$ ; $[z]^{-1} =$ GeV.
}
    \label{fig:HQXzisoT=0mu=0}
\end{figure}

For the heavy quarks model, 
 the behavior of $X(z)$ given by the equation \eqref{X-z} is presented in Fig.\,\ref{fig:HQXzisoT=0mu=0}A, and $\cX=\cX(\varphi)$    is presented in Fig.\,\ref{fig:HQXzisoT=0mu=0}B with different boundary conditions $\varphi_0=-10$ (khaki line) and $\varphi_0=0$ (olive line). The Fig.\,\ref{fig:HQXzisoT=0mu=0}B is recovered from   (\ref{X-z}), (\ref{phiHQ}) and corresponds to exact solutions of the heavy quarks model. The behavior of the dynamical variable $\cal{X}(\varphi)$ is the same as for the light quarks model when $z\to0$.\\

\begin{figure}[h!]  
\centering
\hspace{5em} \includegraphics[scale=0.44]{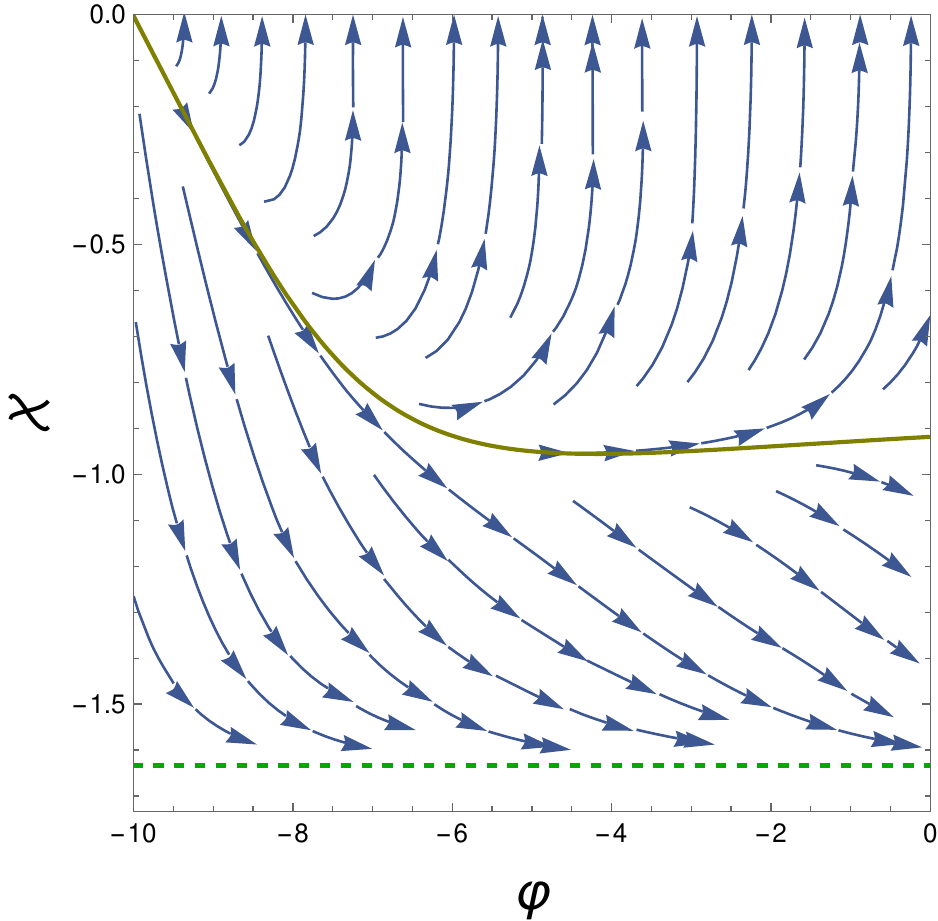}\quad
\includegraphics[scale=0.45]{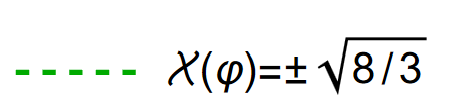}
 \\
  \caption{RG flows for the heavy quarks in isotropic background with $T=0$ and $\mu=0$ corresponding to $\cV(\varphi)$ with boundary condition $\varphi_0=-10$. The green dashed lines represent $\cX=- \sqrt{8/3}$ fixed (attractor) line. The olive line shows the function $\cX$ presented in Fig.\,\ref{fig:HQXzisoT=0mu=0}B.
 }
  \label{Fig:RGHisomu=0T=0}
\end{figure}
 
RG flows for the heavy quarks in the isotropic background at $T=0$ and $\mu=0$ corresponding to $\cV(\varphi)$, (\ref{VH10}), with the boundary condition $\varphi_0=-10$ is depicted in Fig.\,\ref{Fig:RGHisomu=0T=0}.  The approximation for the potential is valid only for $-23.5 \leq \varphi \leq 3.5$  for $\varphi_0=-10$, see Fig.\,\ref{Fig:VPhiHisomu=0T=0}B. The green dashed line represents a fixed (attractor) line  $\cX=- \sqrt{8/3}$. The olive line describes an exact solution of $\cX(\varphi)$ presented in Fig.\,\ref{fig:HQXzisoT=0mu=0}B and corresponds with the repulsive behavior of 
other solutions. It means our solution for the heavy quarks model is unstable.

\newpage

\subsection{RG flow equations for the  non-zero temperature and non-zero chemical potential}

Introducing new dynamical variables
\bea\label{Xz}
X&=&\frac{\dot{\varphi}}{3}\frac{B}{\dot{B}},\\\label{Yz}
Y&=&\frac{1}{4}\frac{\dot{g}}{g}\frac{B}{\dot{B}},\\\label{Hz}
H&=&\frac{\dot{A_t}}{B^2
}
\eea
and using the EOMs (\ref{phi2prime})-(\ref{A2primes}), one obtains the following RG flow equations \cite{Arefeva:2020aan, Arefeva:2018hyo}
\bea\label{cXphi-T}
\frac{d\cX}{d\varphi}&=&-\frac{4}{3} \left(1-\frac{3 }{8}\cX^2+\cY\right) \left(1+\frac{1}{\cX}\frac{2\partial_{\varphi}\cV(\varphi )-\cH^2\partial_{\varphi}\ff_0}{2 \cV(\varphi )+\cH^2\ff_0}\right),\\
\label{cYphi}
\frac{d\cY}{d\varphi}&=&-\frac{4\cY}{3\cX}\left(1-\frac{3 }{8}\cX^2+\cY\right) \left(1+\frac{3}{2\cY}\frac{\cH^2\ff_0}{2\cV(\varphi)+\cH^2\ff_0}\right),\\
\label{cHphi}
\frac{d\cH}{d\varphi}&=& -\left(\frac{1}{\cX}+\frac{\partial_{\varphi}\ff_0}{\ff_0}\right)\cH,
\eea
$\cX(\varphi)$, $\cY(\varphi)$ and $H=\cH(\varphi)$ are related with 
$X$, $Y$ and $H$ as $X=\cX(\varphi(z))$, $Y=\cY(\varphi(z))$ and $H=\cH(\varphi(z))$. We introduced the approximations of $\cV(\varphi)$ in (\ref{VL}), (\ref{VL2}), and for $\ff_0(\varphi)$ in (\ref{fapL}) for the light quarks model. For the heavy quarks, the approximations of the $\cV(\varphi)$ are (\ref{VH10}) and (\ref{VH0}),   and for $\ff_0(\varphi)$ is (\ref{fapH}). The variables $Y$ and $H$ in (\ref{Yz}), (\ref{Hz}) include the temperature and the chemical potential dependence, respectively.
We can also find solutions to the system of the first order differential equations \eqref{cXphi-T}-\eqref{cHphi} which contain more solutions compared to the solutions determined by the potential reconstruction method \cite{Yang:2015aia,
  Arefeva:2018hyo, Arefeva:2022avn}.
 Substituting expressions from \eqref{spsol}-\eqref{g} to \eqref{Xz}-\eqref{Hz},  one can present a solution  for fixed parameters (fixed chemical potential and temperature) for the light and heavy quarks models in  the 3D plots in $(\cX,\cY,\cH)$ coordinates.

\subsubsection{RG flow for the light quarks model}
{\bf a) RG flow for our solution}\\

\begin{figure}[h!]
  \centering
\begin{minipage}{7cm}
    \includegraphics[scale=0.48]{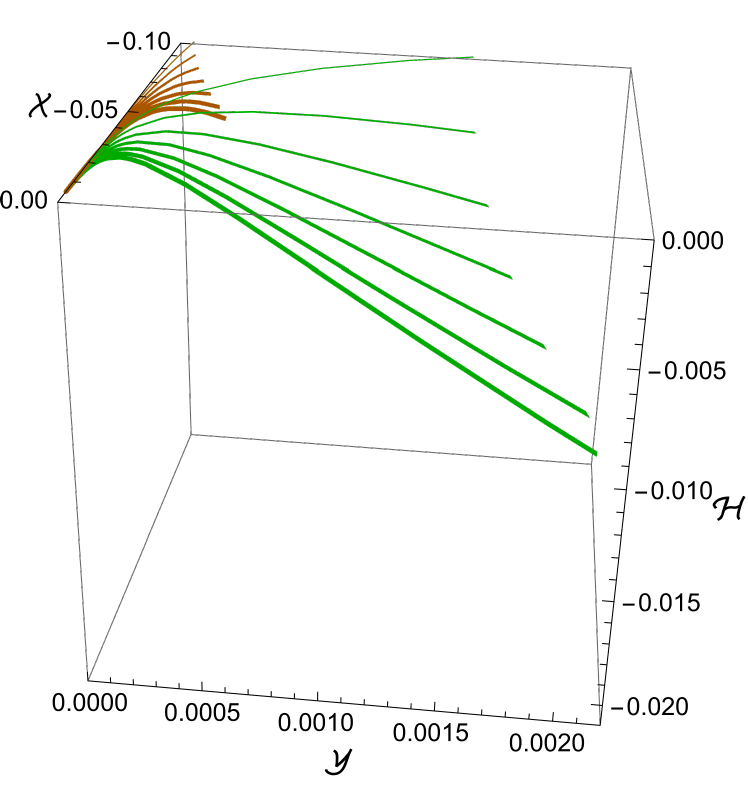} 
  \end{minipage}
  \begin{minipage}{7cm}
\includegraphics[scale=.37]{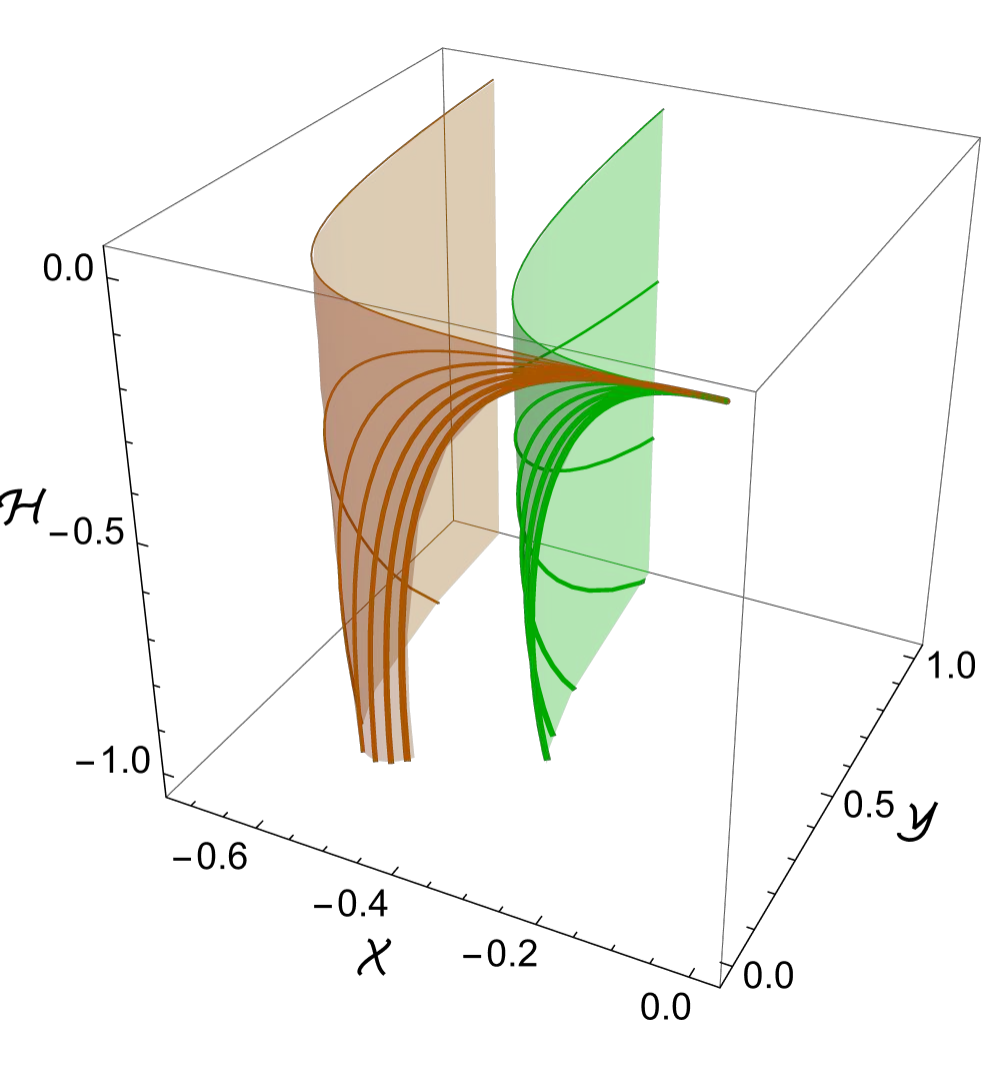}
  \end{minipage}  \\
  \includegraphics[scale=0.10]{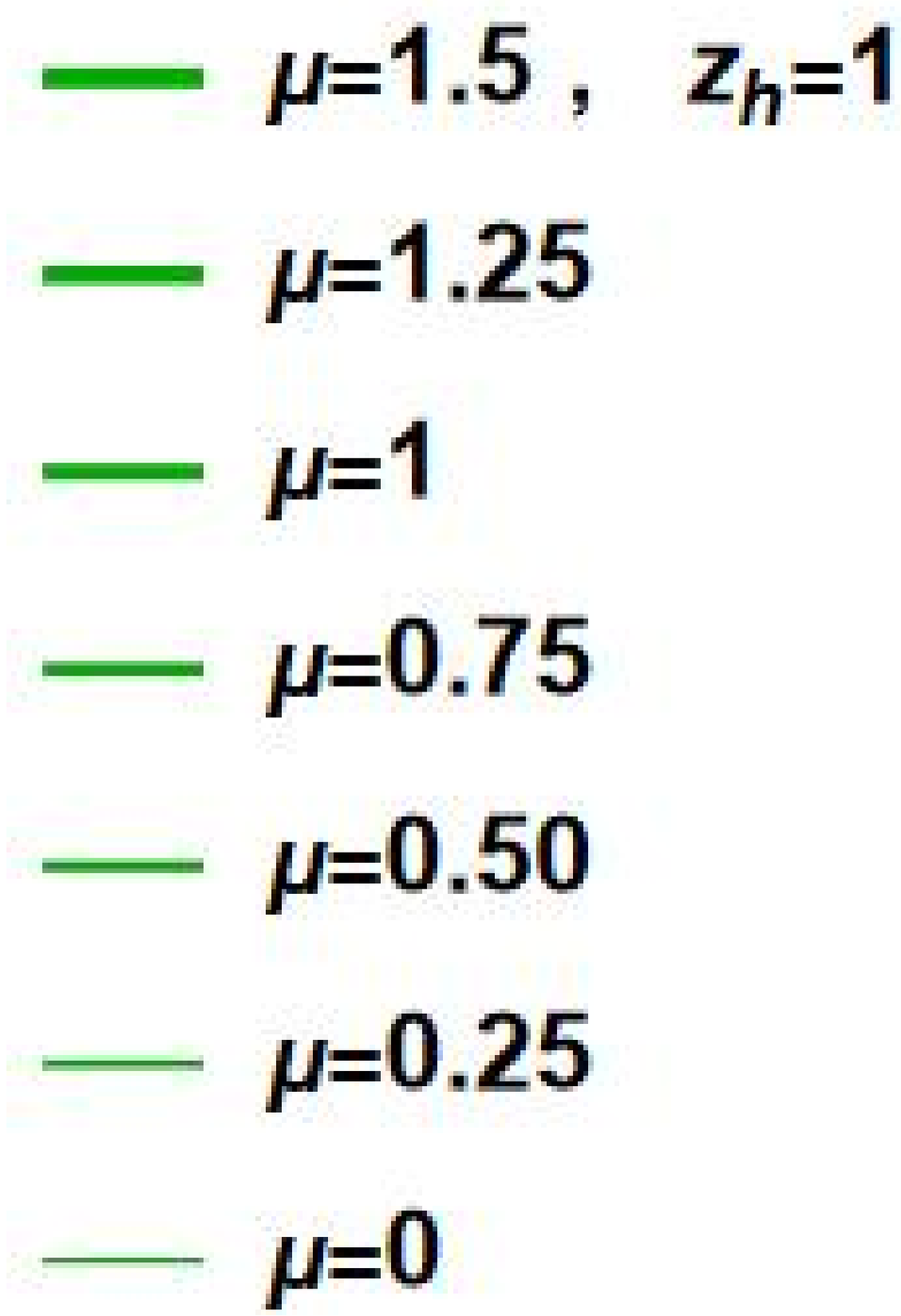}\qquad
  \includegraphics[scale=0.10]{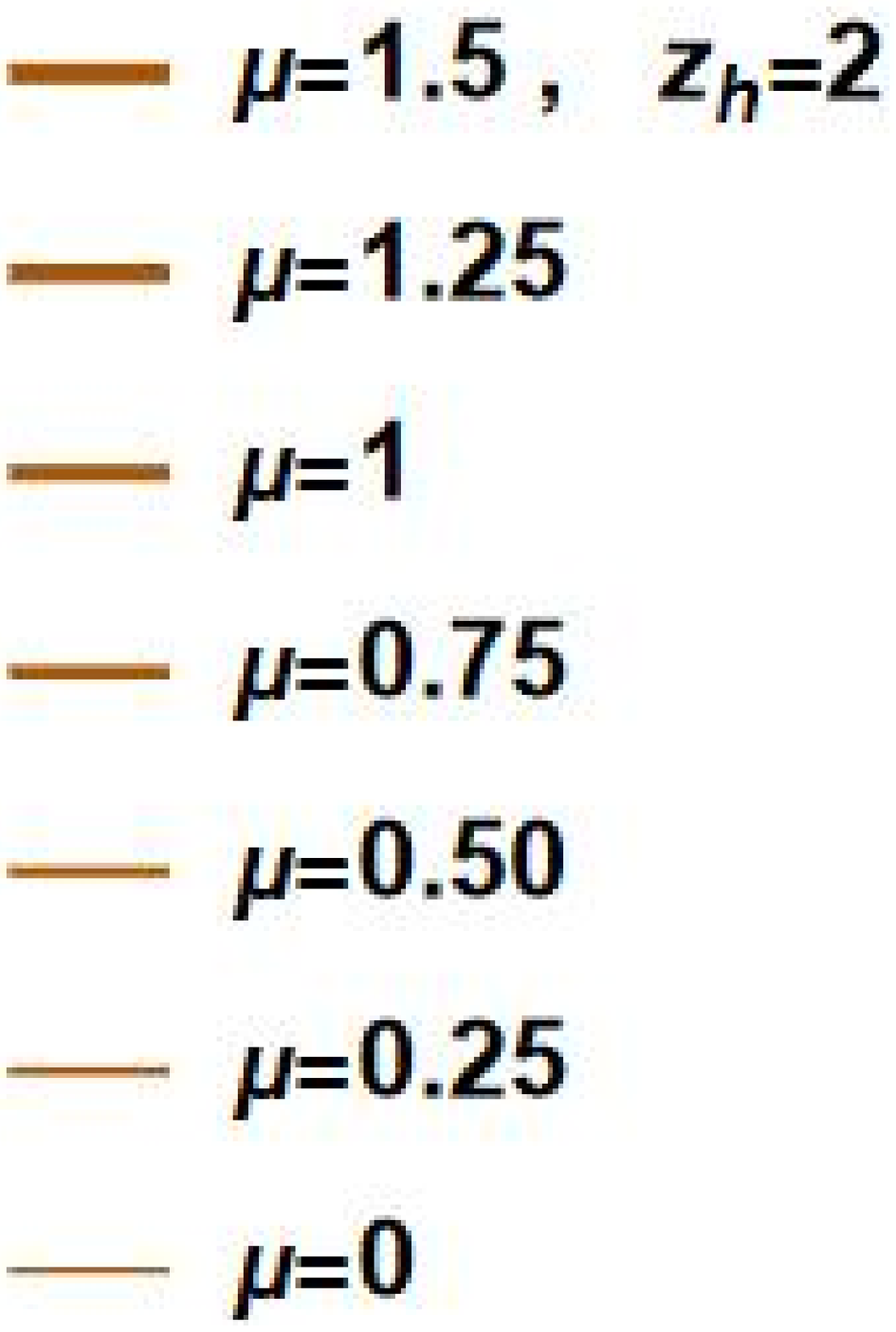}
 \caption{ 
 The 3D plot $(\cX,\cY,\cH)$ for the light quarks model. The green and orange lines represent the solutions with $z_h=1$ and  $z_h=2$ for different $\mu$, respectively; $[\mu]=[z_h]^{-1} =$ GeV. 
 }\label{fig:RGL-3D-lines-phone}
\end{figure}

The 3D plot of $(\cX,\cY,\cH)$ is defined by equations \eqref{Xz}-\eqref{Hz}  for the light quarks model and depicted in Fig.\,\ref{fig:RGL-3D-lines-phone}. The green lines with  decreasing thickness represent the case with $z_h=1$ GeV${}^{-1}$ and $\mu=1.5, 1.25, 1, 0.75, 0.5, 0$ (GeV). The darker orange lines represent the case with $z_h=2$ GeV${}^{-1}$ and the same set of $\mu$. We see that for small values of $z$ brown and green lines almost coincides. This happens because  ${\cal X}(\varphi) \to 0$ for $\varphi \to \varphi_0 $ as in previous case for the 
 zero temperature and chemical potential and, in addition, because of \eqref{X-z}, $\dot{\varphi} \to 0$ and $ B/\dot{B} \to 0$ for $z \to 0$. Also, ${\cal Y}$ and ${\cal H}$ $\to 0$ for $z \to 0$.

\newpage

{\bf b) RG flow for other solutions}\\

\begin{figure}[h!]
  \centering
\includegraphics[scale=0.36]{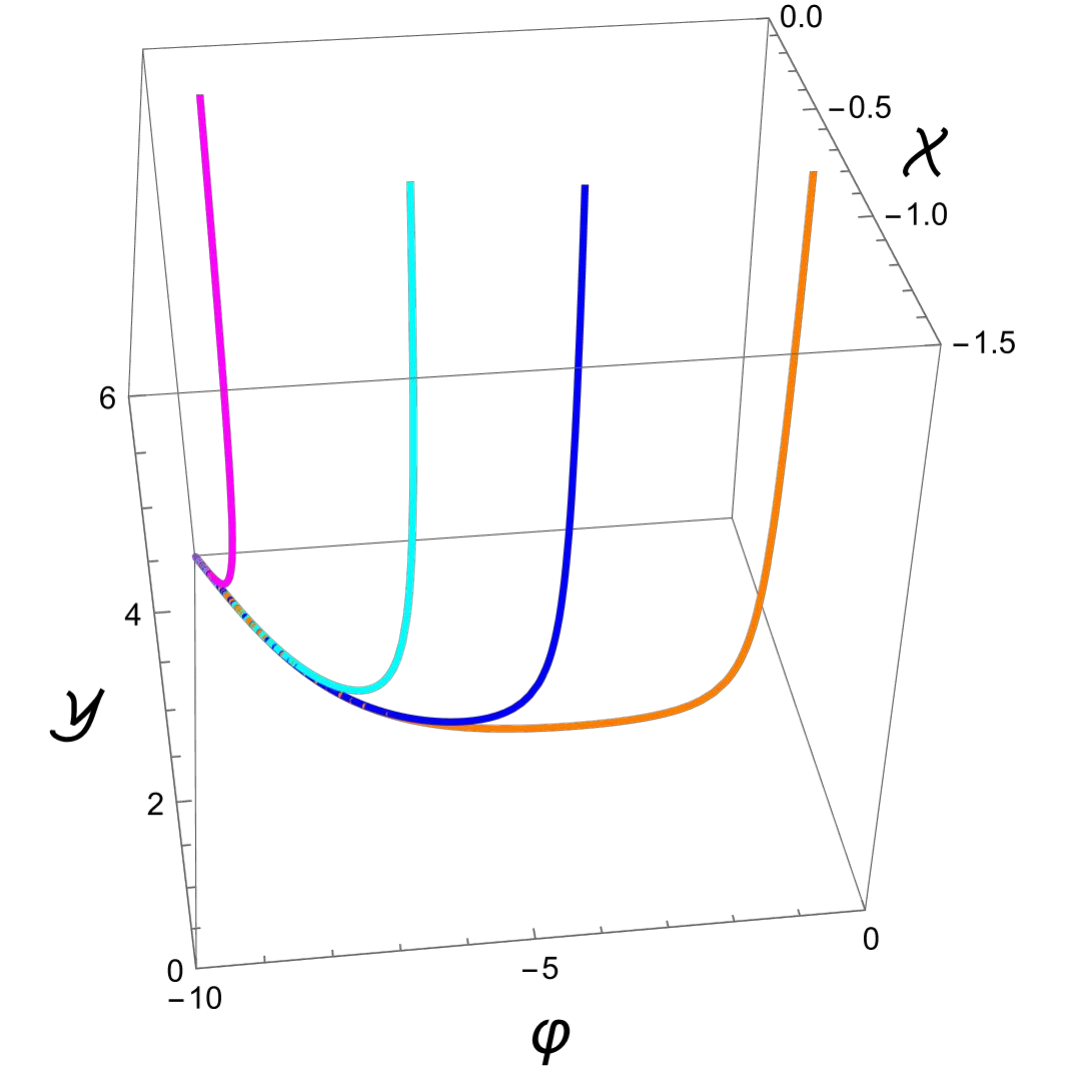}\qquad
\includegraphics[scale=0.50]{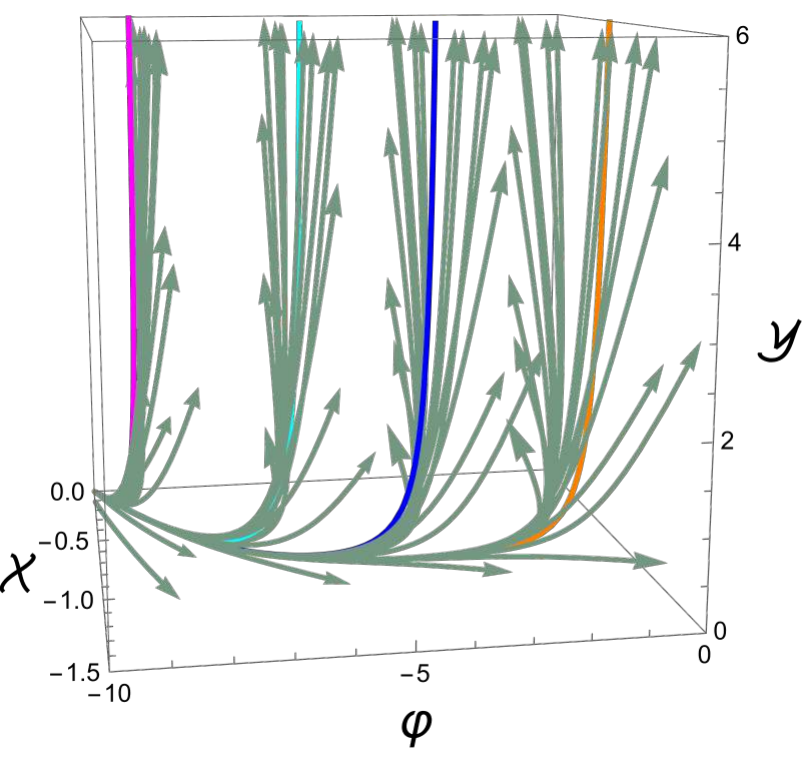}
\hspace{150pt}
A\hspace{150pt}B
\caption{ The 3D plots of $(\cX,\cY,\varphi)$ for the light quarks model representing dependence of $\cal X$ and $\cal Y$  on the dilaton field $\varphi$ with the boundary condition  $\varphi_0=-10$. Different solutions corresponding to different values of  $z_h=0.6, 2.5, 4$, and $6$ are shown by  magenta, cyan, blue and orange, respectively (A), and combination of our solutions in the left panel with 3D RG flows of $(\cX,\cY,\varphi)$ (B); $[z_h]^{-1} =$ GeV. 
}
 \label{Fig:chrisLQ}
\end{figure}

In Fig.\,\ref{Fig:chrisLQ}A, the 3D plot of $(\cX,\cY,\varphi)$ for the light quarks model represents a dependence of $\cal X$ and $\cal Y$ at the zero chemical potential (i.e. ${\cal H}=0$) on the dilaton field $\varphi$ with the boundary condition  $\varphi_0=-10$. Different solutions corresponding to different values of  $z_h=0.6, 2.5, 4$, $6$ (GeV${}^{-1}$) and are shown by  magenta, cyan, blue and orange curves, respectively. These trajectories are obtained for solutions constructed from 
 the potential reconstruction method and the combination of them with 3D RG flows of $(\cX,\cY,\varphi)$ is shown in Fig.\,\ref{Fig:chrisLQ}B. Note that solutions obtained via the  potential reconstruction method are unstable for the light quarks model. For the fixed dilaton potential ${\cal V} (\varphi)$ we solve  \eqref{cXphi-T} and \eqref{cYphi} and obtain different solutions that correspond to different factors $\fb(z)$, $\varphi(z)$ and  ${ V} (z)$ but the same  ${\cal V} (\varphi)$, (\ref{VL}). When we use the reconstruction potential method, we have fixed $\fb(z)$, $f_0(z)$ and fixed boundary conditions for the dilaton field $\varphi$ and blackening function $g$, and obtain only one solution of the EOMs. Different solutions correspond to different boundary conditions of the dilaton field $\varphi$. Therefore, this means that the obtained solution is sensitive to the change of $\fb(z)$. 

\newpage

\subsubsection{RG flow for the heavy quarks model}
\label{RGFHQ}

{\bf a) RG flow for our solution}

\begin{figure}[h!]
  \centering \includegraphics[scale=0.44]{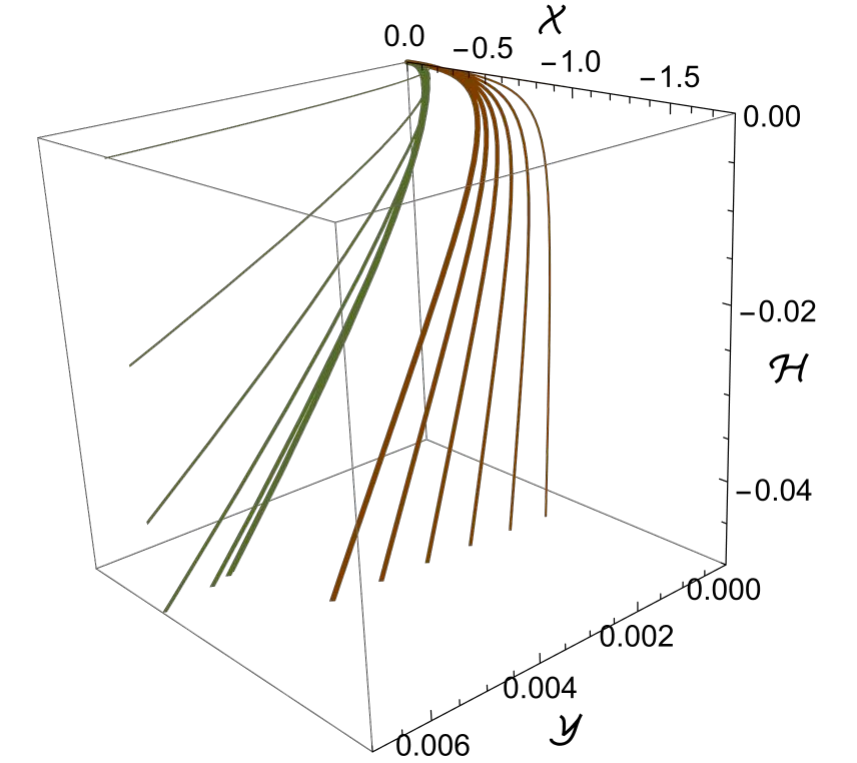} \qquad \includegraphics[scale=0.43]{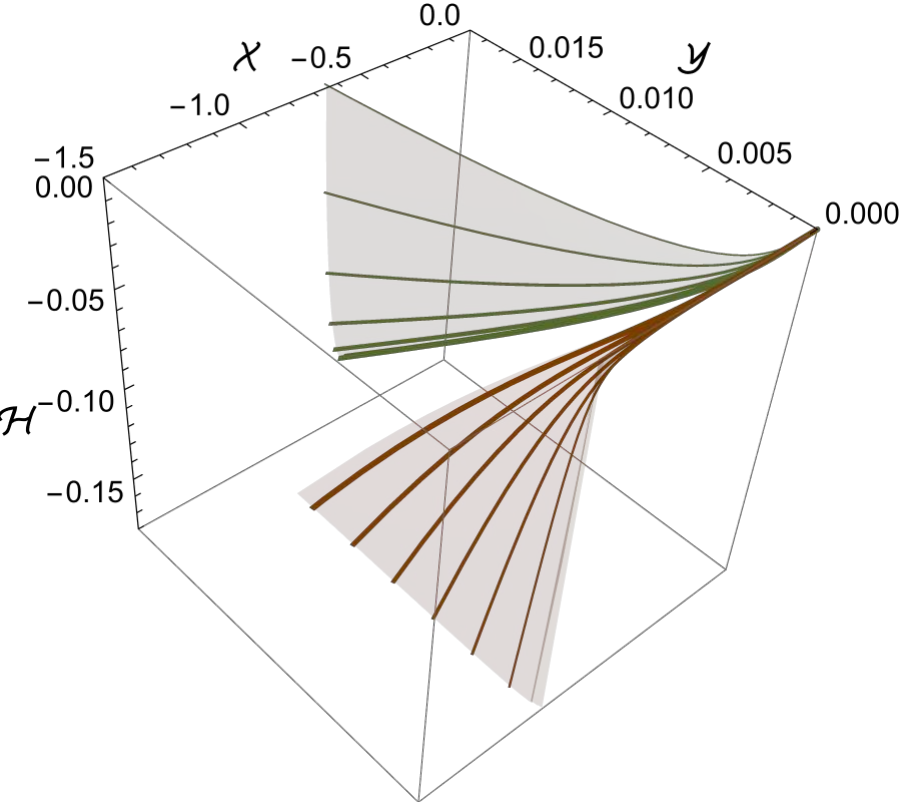}\\
\qquad
\includegraphics[scale=0.40]{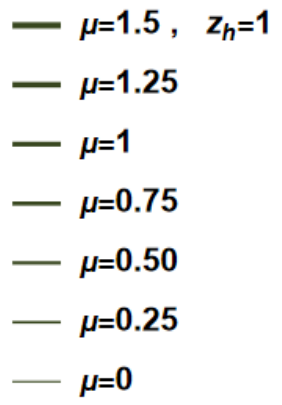}\quad
\includegraphics[scale=0.40]{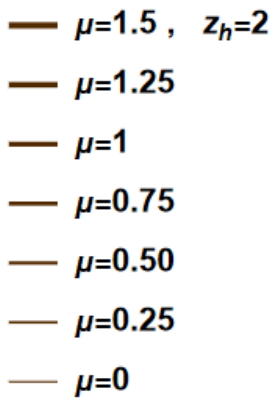}
 \caption{The 3D plot of $(\cX,\cY,\cH)$   for the heavy quarks model. The olive and chocolate lines represent the cases with $z_h=1$ and $z_h=2$ for different $\mu$, respectively; $[\mu]=[z_h]^{-1} =$ GeV. \\
}\label{fig:3DH}
\end{figure}

For the heavy quarks model,  the 3D plot of $(\cX,\cY,\cH)$ defined by equations \eqref{Xz}-\eqref{Hz}   is shown in Fig.\,\ref{fig:3DH}. The olive lines with decreasing thickness represent the case with $z_h=1$ GeV${}^{-1}$ and $\mu=1.5, 1.25, 1, 0.75, 0.5, 0$ (GeV). The chocolate lines represent the case with $z_h=2$ GeV${}^{-1}$ and the same set of $\mu$. We can see the similar behaviour for the asymptotic as for the light quarks case, i.e. ${\cal X}(\varphi) \to 0$ for $\varphi \to \varphi_0$. 

\newpage

{\bf b) RG and other solutions}\\

\begin{figure}[h!]
  \centering
\includegraphics[scale=0.43]{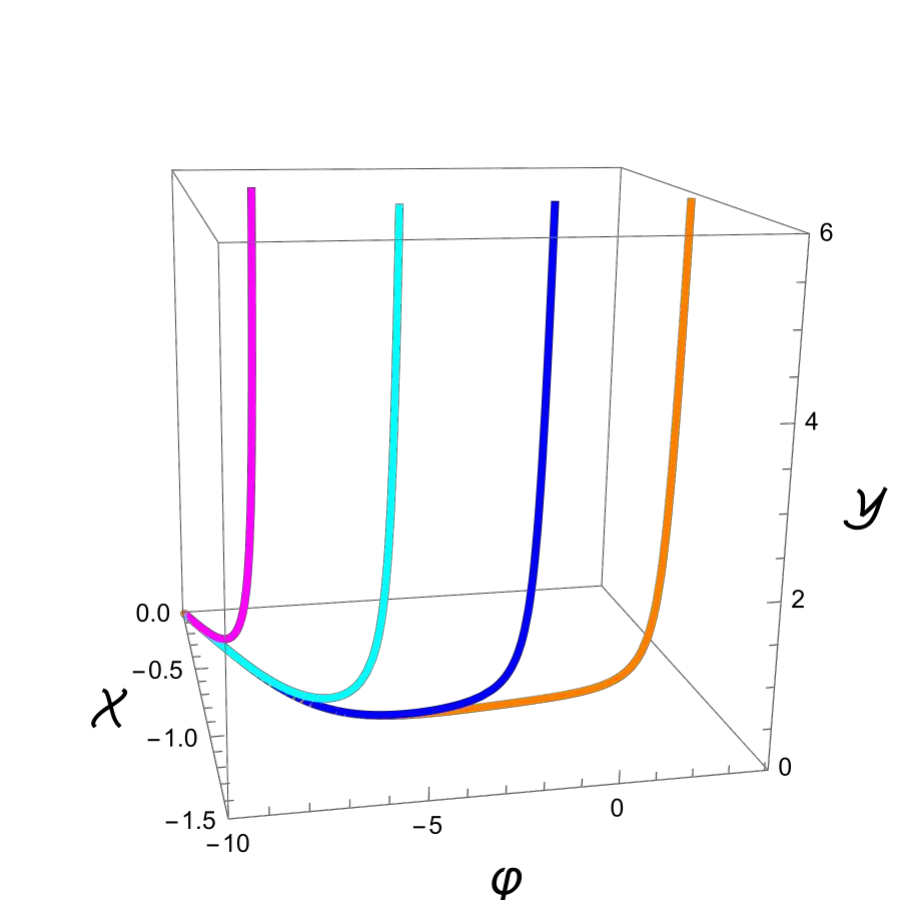}\qquad
\includegraphics[scale=0.44]{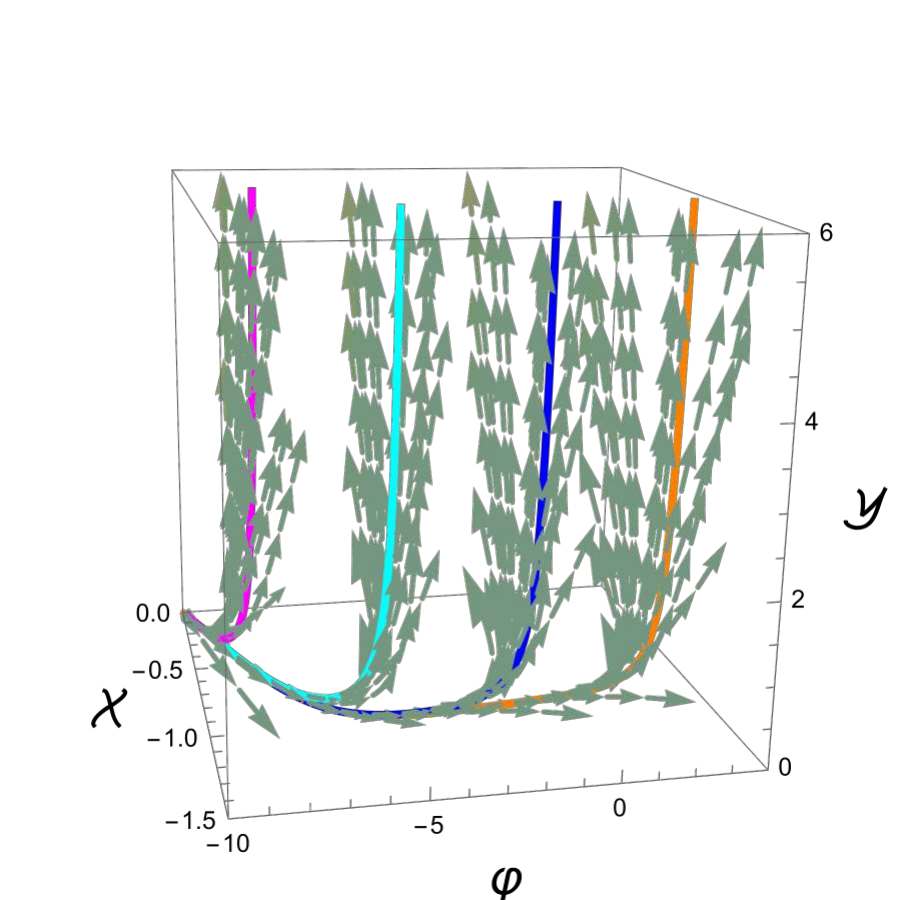}\\
A\hspace{200pt}B\\
\caption{The 3D plots of $(\cX,\cY,\varphi)$ for the heavy quarks model representing dependence of $\cal X$ and $\cal Y$  on the dilaton field $\varphi$ with the boundary condition  ${\varphi_0}=-10$. Different solutions correspond to different values of $z_h=0.45, 1.0, 1.4$, $1.65$ are shown by  magenta, cyan, blue and orange, respectively (A), and combination of our solutions in the  left panel with 3D RG flows of $(\cX,\cY,\varphi)$ (B); $[z_h]^{-1} =$ GeV.}
 \label{Fig:HQstreamsol}
\end{figure}

The 3D plots of $(\cX,\cY,\varphi)$ for the heavy quarks model representing dependence of $\cal X$ and $\cal Y$  on the dilaton field $\varphi$ at the zero chemical potential (i.e. ${\cal H}=0$) with the boundary condition  ${\varphi_0}=-10$ is shown in Fig.\,\ref{Fig:HQstreamsol}A. Different solutions correspond to different values of $z_h=0.45, 1.0, 1.4$, $1.65$ (GeV${}^{-1}$) are shown by  magenta, cyan, blue and orange, respectively. These solutions are combined with 3D RG flows of $(\cX,\cY,\varphi)$, i.e. other solutions of RG equations for the same dilaton potential ${\cal V}(\varphi)$, (\ref{VH10}), are shown  in Fig.\,\ref{Fig:HQstreamsol}B. The solutions obtained via the potential reconstruction method are unstable for the heavy quarks the same as for the light quarks case. For the fixed dilaton potential ${\cal V} (\varphi)$ we solve  \eqref{cXphi-T} and \eqref{cYphi} and obtain different solutions that correspond to different factors $\fb(z)$, $\varphi(z)$ and  $V(z)$ but the same  ${\cal V} (\varphi)$.
When we use the reconstruction potential method we have fixed $\fb(z)$, $f_0(z)$ and fixed boundary conditions for the dilaton field $\varphi$ and blackening function $g$, and obtain only one solution of the EOMs. Therefore, this means that the obtained solution is sensitive to the change of $\fb(z)$. 
\\

\newpage

\section{Conclusion and Discussion} \label{sec:concl}
In this paper, we considered the $\beta$-function in  holographic models supported by Einstein-dilaton-Maxwell action for  heavy and light quarks.  We obtained a significant dependence of the  $\beta$-function on mass of quarks, chemical potential and temperature. 
 At the 1st order phase transitions, $\beta$-functions undergo jumps depending on temperature and chemical potential. For the heavy quarks these jumps depend more sharply on the thermodynamic parameters. 
Note that a choice of  a boundary condition has a crucial effect on the dilaton field. For this reason we chose the second boundary condition $z_0=\fz(z_h)$, i.e. \eqref{phi-fz-LQ} for the light quarks and \eqref{bceHQ} for the heavy quarks, to respect the lattice calculations for QCD string tension as a function of temperature $\sigma(T)$  for the non-zero temperature and zero chemical potential \cite{Kaczmarek:2007pb}. Therefore, the second boundary condition is a physical one.  Although the first boundary condition is not physical the $\beta$-function can  exhibit a jump from hadronic to quarkyonic phases. 
\\

To summarize our findings:

\begin{itemize}

\item The dependence of the $\beta$-function for the light quarks model on the thermodynamic parameter $T$ is shown in Sect.\,\ref{LQ-nbc10}. We see that
\begin{itemize}
\item in all regions the $\beta$-function is negative;

\item in the region of the hadronic phase, the $\beta$-function increases very fast with increasing temperature, at fixed $\mu$ and energy scale $E$;

\item in the region of the QGP phase, the $\beta$-function increases much more slower with increasing temperature in comparison to the hadronic phase, at fixed $\mu$ and energy scale $E$;
\item on the 1st order transition line, the $\beta$-function has a jump; the magnitude of the jump is 
zero at the CEP. The magnitude of the jump increases by decreasing the probe energy scale $E$;

\end{itemize}

 \item The dependence of the $\beta$-function for the heavy quarks model on the thermodynamic parameter $T$ is shown in Sect.\,\ref{HQ-nbc15}. We see that
\begin{itemize}
\item in all regions the $\beta$-function is negative;
\item in the region of the hadronic phase, the $\beta$-function increases  slowly with increasing temperature, at fixed $\mu$ and energy scale $E$; 

\item in the region of the QGP phase, the $\beta$-function increases faster with increasing temperature in comparison to the hadronic phase, at fixed $\mu$ and energy scale $E$;

\item on the 1st order transition line, the $\beta$-function has a jump; 
the magnitude of the jump is 
zero at the CEP. The magnitude of the jump increases by decreasing the probe energy scale $E$;

\end{itemize}

\item The behavior of the $\beta$-function in terms of the temperature $T$ is given by the formula $\beta(\alpha)=3 \alpha\, X$.  It happens that $\alpha$  varies with $T$ rather fast in the hadronic phase, see Fig. 25a, 30a, and Fig. 30c (for light quarks) and  Fig. 40, 42a, and Fig. 42c   (for heavy quarks) in our previous paper \cite{AHSU}. This behavior of the running coupling constant is due to the choice of boundary conditions for the dilaton field given by Eq.(2.19) for light quarks and Eq.(2.33) for heavy  quarks in \cite{AHSU}. They  are  the boundary conditions that ensure the temperature dependence of the string tension between quarks in the hadron phases, which are obtained on the lattice.  The factor  $X$  is defined by \eqref{X-z} and does  depend on the derivative of the dilaton field, $\dot \phi$, and does not depend on the dilaton boundary conditions. Let us note the dependence of running coupling on the temperature is not so drastic for heavy quarks in comparison to the light quarks \cite{AHSU}.

\item  At low temperature, i.e. in the hadronic phase, the $\beta$-function becomes very small but it is not exactly zero both for the light and heavy quarks models.

   \item We also studied the RG flow near our solution. Our solution is unstable, that  is inevitably associated with a negative potential, that in turn is associated with the existence of the hadronic phase. 
  \item  The RG flow for the light and heavy quarks is approximately the same although the associated warp factor is completely different.
  \item Note that for the heavy quarks jumps depend more sharply on thermodynamic parameters. The magnitude of the jumps increases with decreasing energy scales for both light and heavy quarks.
  \item 
 The choice of different boundary conditions simply shifts the RG for both heavy and light quarks without leading to significant changes.
\end{itemize}

It is important to note that to obtain the RG flows for heavy and light quarks we utilized some approximations for dilaton potential and the gauge kinetic function. In fact, from equation \eqref{At2prime} we constructed the dilaton field as a function of $z$, then using 
\eqref{A2primes} we get the potential  as a function of $z$. Then taking into account that the dilaton function is monotonic, we reconstructed $\cV(\varphi)$ and  $\ff_0(\varphi)$ as a functions of $\varphi$.
Therefore, we have no direct formula for $\cV(\varphi)$ and  $\ff_0(\varphi)$.
 \\

At the 1st order phase transition, some physical quantities also have jumps in different holographic models, namely, running coupling \cite{AHSU}, entanglement entropy \cite{Dudal:2018ztm,Arefeva:2020uec}, electric conductivity and direct photons emission rate \cite{Arefeva:2022avn, Arefeva:2021jpa} and  energy loss \cite{Arefeva:2020bjk,Arefeva:2021btm}. In addition, the effect of primary (spatial) anisotropy on the QCD phase transition is studied in \cite{Arefeva:2018hyo,Arefeva:2018cli}, and the effect of the magnetic field is considered in \cite{Gursoy:2017wzz,Bohra:2019ebj,Bohra:2020qom,Arefeva:2020vae, Dudal:2021jav,Jain:2022hxl,Arefeva:2022avn}. The effect of the magnetic field on running coupling is considered in \cite{Arefeva:2024xmg,AHNS}.
It would be interesting to investigate these new parameters on the $\beta$-function in future research. Also, $\beta$-function as a function of running coupling is studied in another paper \cite{Arefeva:2024poq}.
\\

\section{Acknowledgments}

The work of I. A. and M. U. is supported by Theoretical Physics and Mathematics Advancement Foundation ``BASIS” (No. 24-1-1-82-1). The work of P. S. is supported by Theoretical Physics and Mathematics Advancement Foundation ``BASIS” (No. 23-1-4-43-1). The work of A. H. was performed at the Steklov International Mathematical Center and supported by the Ministry of Science and Higher Education of the Russian Federation (Agreement No. 075-15-2022-265).


\newpage 
\appendix
\appendix

\section{EOM for the light and heavy quarks models} \label{appendix A}

Varying the action (\ref{action}) and applying the anzatz (\ref{metric})-(\ref{warp-factor}), one obtains the EOMs in the following form \cite{Li:2017tdz,Yang:2015aia,
  Arefeva:2018hyo, Arefeva:2022avn, Arefeva:2022bhx, Arefeva:2020vae,Arefeva:2020byn,Arefeva:2021mag,Arefeva:2023ter} 
\bea
\label{phi2prime}
\varphi''+\left(\frac{g'}{g}+3A'-\frac{3}{z}\right)\varphi'+\left(\frac{z^2e^{-2A}A'_tf_{0,\varphi}}{2g}-\frac{e^{2A}V_{\varphi}}{z^2g}\right)=0,\\\label{At2prime}
A''_t+\left(\frac{f'_{0}}{f_{0}}+A'-\frac{1}{z}\right)A'_t=0,\\
\label{phiprime}
A''-A'^2+\frac{2}{z}A'+\frac{\varphi'^2}{6}=0,\\\label{g2prime}
g''+\left(3A'-\frac{3}{z}\right)g'-e^{-2A}z^2f_0 A_t'^2=0,\\\label{A2primes}
A''+3A'^2+\left(\frac{3g'}{2g}-\frac{6}{z}\right)A'-\frac{1}{z}\left(\frac{3g'}{2g}-\frac{4}{z}\right)+\frac{g''}{6g}+\frac{e^{2A}V}{3z^2g}=0.
\eea
here all functions  depend on the omitted holographic coordinate $z$   and $V(z)=\cV(\varphi(z))$,  
$V_\varphi(z)=\cV_\varphi(\varphi(z))$ and $f_0(z)=\ff_0(\varphi(z))$. 
\\

The solutions to the EOMs \eqref{phi2prime}-\eqref{A2primes} are:
\bea
\label{phiprime} 
\varphi'(z)&=&\sqrt{-6\Bigg(A''-A'^2+\frac{2}{z}A'\Bigg)},\label{spsol}\\
A_t(z)&=&\mu\frac{e^{cz^2}-e^{cz^2_h}}{1-e^{cz^2_h}},
\\
g(z)&=&1-\frac{1}{\int_0^{z_h}y^3e^{-3A}dy}\Bigg[\int_0^zy^3 e^{-3A}dy-\frac{2c\mu^2}{(1-e^{cz_h^2})^2}\Bigg|\begin{matrix}
\int_0^{z_h}y^3e^{-3A}dy & \int_0^{z_h}y^3e^{-3A}e^{cy^2}dy \\
\int_{z_h}^z y^3e^{-3A}dy & \int_{z_h}^{z}y^3e^{-3A}e^{cy^2}dy
\end{matrix}\Bigg|\Bigg],\nonumber\\
\label{g}
\\\nn
\,
\\
V(z)&=&-3z^2g e^{-2A}\Bigg[A''+3A'^2+\Bigg(\frac{3g'}{2g}-\frac{6}{z}\Bigg)A'-\frac{1}{z}\Bigg(\frac{3g'}{2g}-\frac{4}{z}\Bigg)+\frac{g''}{6g}\Bigg].\label{Vsol}
\eea

\subsection{Preliminary about the reconstruction method} \label{Reconst-method}

Let us consider the following action
\bea
S&=&\frac{1}{16\pi G_5}\int d^5x\sqrt{-\fg} \left[R-\frac{1}{2}\partial_{\mu}\varphi\partial^{\mu}\varphi-\cV(\varphi)\right].\label{Raction}
\eea 
For zero temperature, the blackening function  $g(z)=1$, the metric (\ref{metric}) can be written as
\be
ds^2=\frac{e^{2A(z)}}{z^2}\left[-dt^2+d\Vec{x}^2+dz^2\right].\label{mT0}
\ee  
 Then for $T=0$ and $\mu=0$, the Einstein equations that follow from the action \eqref{Raction} for the ansatz for metric \eqref{mT0} are given by 
\bea
\label{phi2primeT0mu0}
\varphi''+\left(3A'-\frac{3}{z}\right)\varphi'-\frac{e^{2A}V_{\varphi}}{z^2}=0,\\
\label{A2primeT0mu0}
A''-A'^2+\frac{2}{z}A'+\frac{\varphi'^2}{6}=0,\\
\label{VT0mu0}
A''+3A'^2-\frac{6}{z}A'+\frac{4}{z^2}+\frac{e^{2A}V}{3z^2}=0.
\eea

From the equation \eqref{A2primeT0mu0} we get the expression for the  dilaton field
\bea
\varphi'(z)&=&\sqrt{-6\Bigg(A''-A'^2+\frac{2}{z}A'\Bigg)},\label{Rspsol}
\eea
and from the equation \eqref{VT0mu0} we get the expression for the  potential depending on $z$ coordinate
\bea
V(z)&=&-3z^2 e^{-2A}\Bigg[A''+3A'^2+\Bigg(-\frac{6}{z}\Bigg)A'-\frac{1}{z}\Bigg(-\frac{4}{z}\Bigg)\Bigg].\label{RVsol}
\eea
For the boundary condition $\varphi(z_0)=0$ we define
\bea
\varphi(z,z_0)&=&\int _{z_0}^z \sqrt{-6\Bigg(A''-A'^2+\frac{2}{z}A'\Bigg)}dz \label{phians}\eea
So, our solution depends on $z_0$.
We can denote our solution as 
$\varphi_{z_0}(z)$.
We define
\be
\cV(\varphi,z_0)=
V\left(\cZ(\varphi,z_0)\right)
\ee
where $V(z)$ follows from \eqref{RVsol} and $\cZ$ is the inverse to  the function $\varphi_{z_0}=\varphi_{z_0}(z)$
\be
\cZ: \varphi\to z,\qquad \mbox{such that} \qquad \varphi_{z_0}(\cZ(\varphi_{z_0}))=\varphi_{z_0}
\ee
We can draw the form of $\cV=\cV(\varphi_{z_0})$ using the parametric plot.

Introducing a new variable (\ref{X-z}), one obtains the RG equation (\ref{cXphi}). Now we solve this equation  with a boundary condition 
\be\label{RbcX}
\cX\Big|_{\varphi=\varphi_1}=\cX_1.\ee
We take $X=X(z)$ and solution $\varphi _{z_0}(z)$ with fixed  $z_0$
and draw 
\be
\cX=\cX (\varphi)\ee
using a parametric plot.
Now  we take some value of $\varphi=\varphi_1$ and in the parametric plot find $\cX_1$
\be
\label{bc}
\cX_1=\cX (\varphi_1)\ee
We can solve  the first order differential equation (\ref{cXphi}) with the  boundary condition \eqref{bc}. Denoting the 
 solution  $\cX=\cX(\varphi,\cX_1,\varphi_1)$, we can find 
$X=X(z)$
as 
\be
X(z)=\cX(\varphi_{z_0}(z),\cX_1,\varphi_1)\ee
the result should be independent of $z_0$ and coincide  
with (\ref{X-z}).

\subsection{Reconstruction of the potential}

\subsubsection{Potential of the light quarks model}

Substituting the scale factor (\ref{wfL}) into (\ref{spsol}) and (\ref{Vsol}), we obtain the solution for the dilaton field as
\bea\label{Lphiz}
\varphi(z,\varphi_0)=\varphi_0 
&+&2 \sqrt{3a}  \left(\!\!\sqrt{2 a+1}\, \arcsinh\left(\sqrt{\frac{b(2
   a+1)}{3}} \, z\right)\right.\\
&-& \left.\sqrt{2(a-1) }\,\arctanh
   \left( \sqrt{\frac{{2b(a-1)} \, }{(2 a+1) b
z^2+3}}\,z\right)\!\!\right)  \nn \eea
and the potential of the dilaton
\bea
   \label{LVz}
V(z)&=&-6 \left(b z^2+1\right)^{2 a-2} \left[b z^2 \left((a (6 a+7)+2) b z^2+5
   a+4\right)+2\right].
\eea

The functions $\varphi(z)$  and  $V(z)$ given by \eqref{Lphiz} and \eqref{LVz} are presented in Fig.\,\ref{Fig:PhiHisomu=0T=0}A
and Fig.\,\ref{Fig:PhiHisomu=0T=0}B, respectively.  The function $\cV(\varphi)$ that we recover from the given functions $V(z)$, eq.\eqref{LVz}, and  $\varphi(z)$ depending on $\varphi_0$, eq.\eqref{Lphiz}, is presented in Fig.\,\ref{Fig:VPhiHisomu=0T=0}A. We see  that $\cV(\varphi)$ depends on $\varphi_0$. \\

When the boundary condition for the dilaton field is selected as $\varphi_0=-10$ at $z=0$, we can approximate the light quarks potential function Fig.\,\ref{Fig:VPhiHisomu=0T=0}A by  
\be\label{VL}
\cV_{approx,-10}(\varphi)=\sum_{i=0}^{18} c_i \varphi^i,
\ee
and by fitting the points of the graph on the domain $-23.5\leq\varphi\leq3.5$, we obtain the following coefficients  that are given in  Table\,\ref{table1} in  Appendix\,\ref{approx1}. \\

With the boundary condition $\varphi_0=0$ at $z=0$, the approximation for the light quarks potential takes the form
\be \label{VL2}
\cV_{approx,0}(\varphi)=\sum_{j=0}^{18}b_j \varphi^j,
\ee
on the domain $-13.5\leq\varphi\leq13.5$ with fitting coefficients that  are given in the Table\,\ref{table1} in  Appendix\,\ref{approx1}. \\

\subsubsection{Potential of the heavy quarks model}

The solution for the dilaton field by inserting the scale factor (\ref{scaleHQ}) into the expression (\ref{phiprime})  as
\bea\nn \label{phiHQ}
\varphi(z,\varphi_0)=&\sqrt{6}& \int_{0} ^{z}
~ d\xi
\sqrt{\left(-4\, p\, \xi^3-\frac{2\, \mathrm{s}\, \xi}{3}\right)^2+2 \left(4\, p\,
   \xi^2+\frac{2\, \mathrm{s}}{3}\right)+12\, p \,\xi^2+\frac{2\, \mathrm{s} }{3}}\\
   &+& \varphi_0
\eea
and the potential of the dilaton
\be\label{HVz}
V(z)=-2 ~ e^{\frac{1}{3}(6\,p\, z^4+2\, \mathrm{s}\, z^2) }\Bigg(6+5\,\mathrm{s} \, z^2 + 2 (9\, p + {\mathrm{s}}^2) z^4 + 24\,p\, \mathrm{s} \, z^6 + 72\, p^2 z^8\Bigg). 
\ee
\begin{figure}[b!]
  \centering
   \includegraphics[scale=0.35]{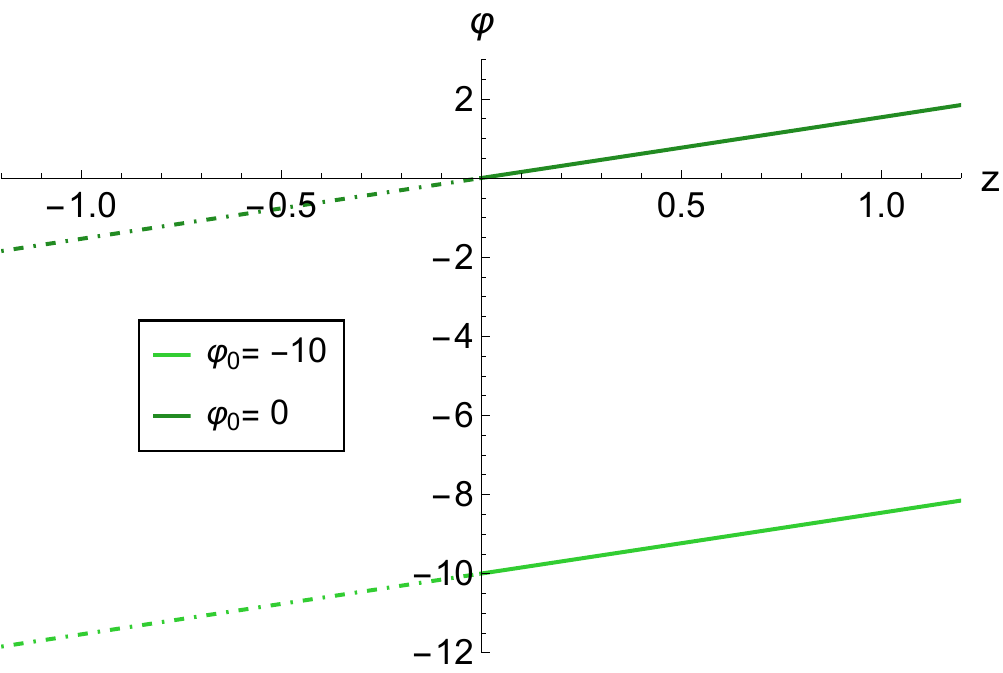} \qquad
\includegraphics[scale=0.35]{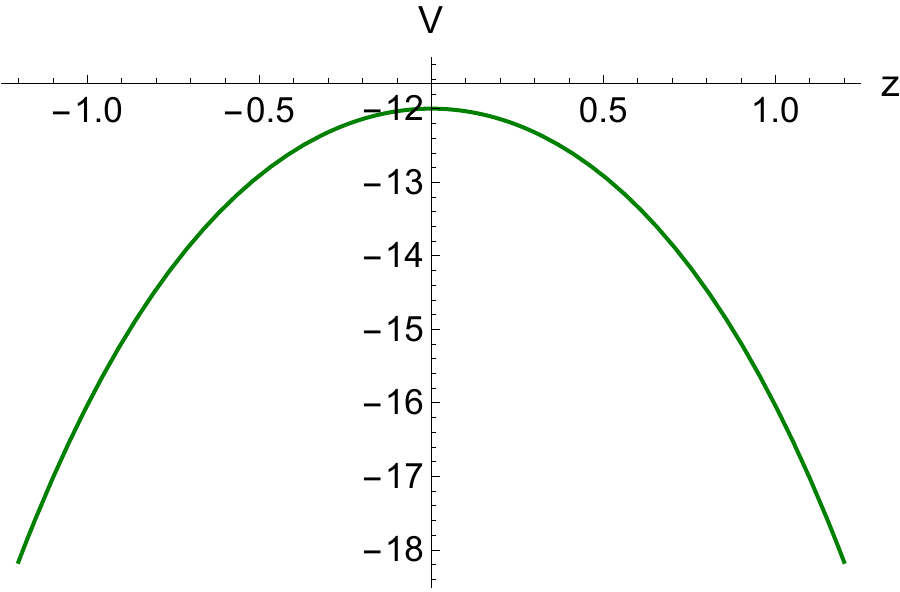} \\
A\hspace{200pt}B\\
  \includegraphics[scale=0.36]{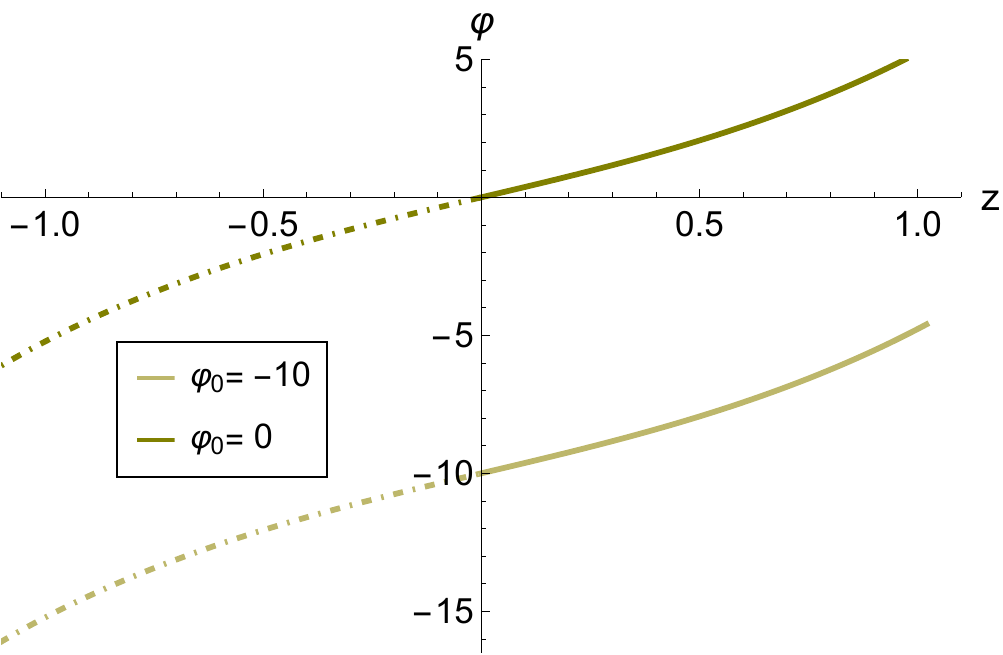} \qquad
\includegraphics[scale=0.35]{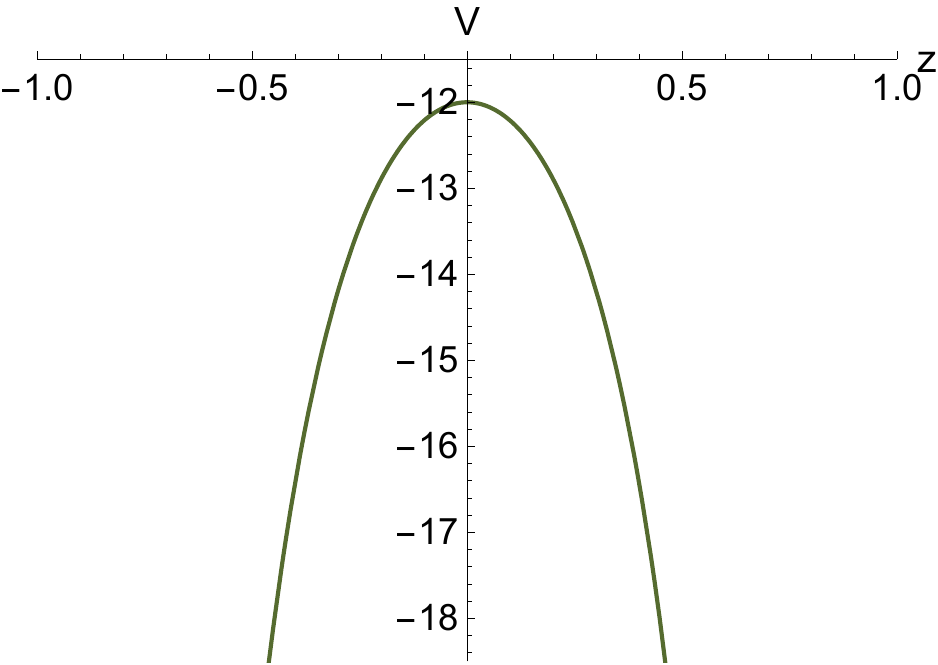} \\
 C \hspace{200pt}D
  \caption{The dilaton field $\varphi$ for the light quarks (A) and heavy quarks (C) as a function of $z$ for the different boundary conditions $\varphi_0=-10$ and $\varphi_0=0$, and the potential $V$ as a function of $z$ for the light quarks (B) and for the heavy quarks (D); $[z]^{-1} =$ GeV. 
  }
\label{Fig:PhiHisomu=0T=0}
\end{figure}

\begin{figure}[h!]
  \centering
  \includegraphics[scale=0.33]{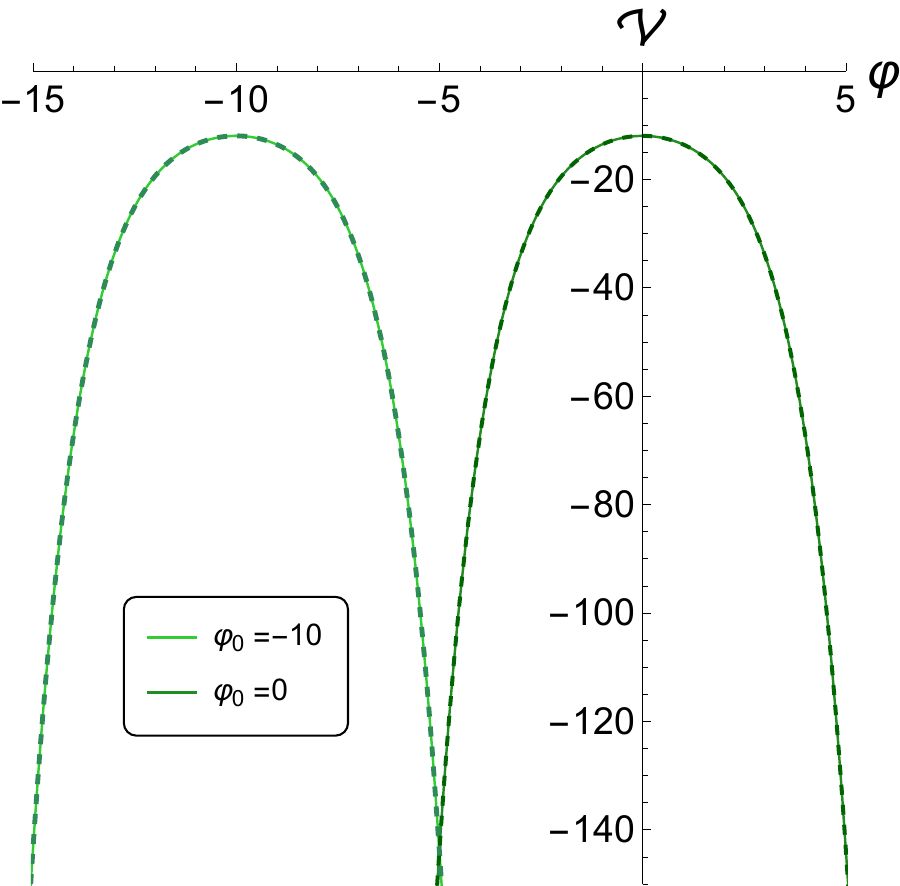} \quad
  \includegraphics[scale=0.32]{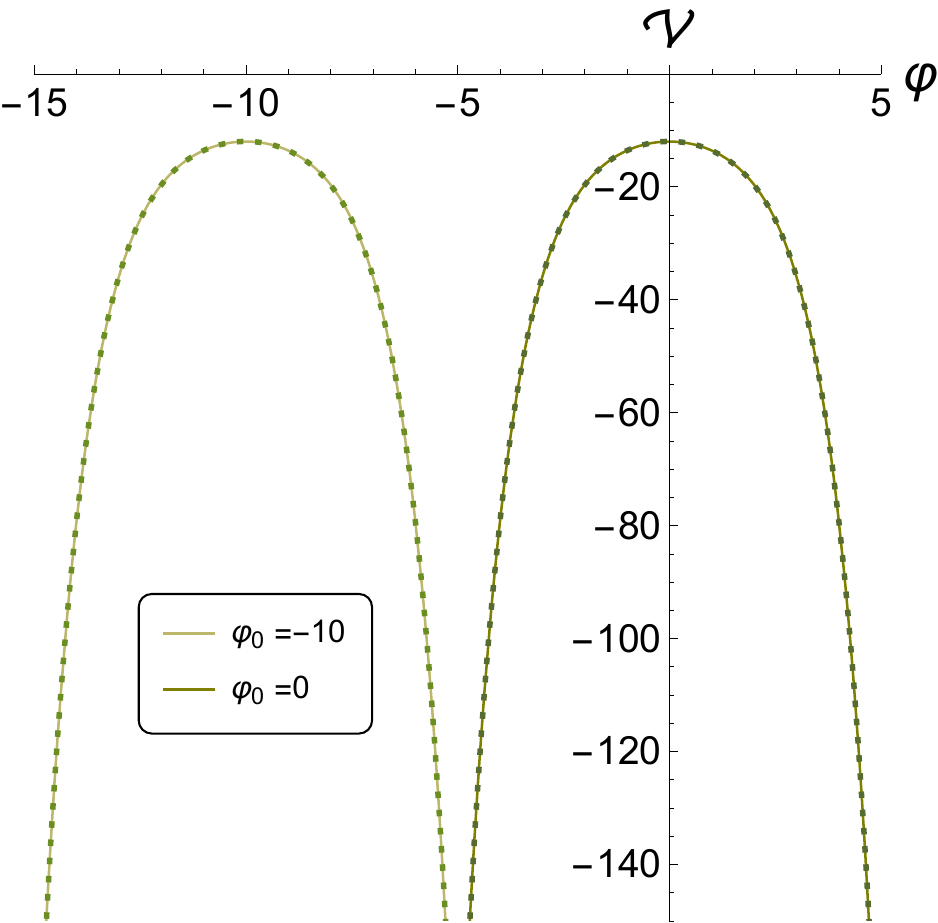} \\
  A\hspace{15em}B
 \caption{The light quarks (A)  and heavy quarks (B) potentials $\cV$ as functions of the dilaton field $\varphi$ with the boundary conditions $\varphi_0=-10$  and $\varphi_0=0$. The dashed lines  correspond to the approximation functions (\ref{VL}) and (\ref{VL2}) for the light quarks and (\ref{VH10}) and (\ref{VH0}) for the heavy quarks, respectively to the boundary condition. \\
}
  \label{Fig:VPhiHisomu=0T=0}
\end{figure}

The functions $\varphi(z)$  and  $V(z)$ given by \eqref{phiHQ} and \eqref{HVz} are shown in Fig.\,\ref{Fig:PhiHisomu=0T=0}C
and Fig.\,\ref{Fig:PhiHisomu=0T=0}D, respectively. The potential of the heavy quarks as a function of the dilaton field  is presented in Fig.\,\ref{Fig:VPhiHisomu=0T=0}B. 
Considering the boundary condition for the dilaton field as $\varphi_0=-10$ at $z=0$, one can approximate the heavy quarks potential function Fig.\,\ref{Fig:VPhiHisomu=0T=0}B (a khaki dashed line) by  
\be \label{VH10}
\cV_{approx,-10}(\varphi)=\sum_{i=0}^{18} \mathfrak{c}_i \varphi^i,
\ee
and by fitting the points of the graph on the domain $-23.5\leq\varphi\leq3.5$, we obtain the following coefficients  that  are given in  Table\,\ref{table2} in  Appendix\,\ref{approx1}. 
Also, for the choosing boundary condition as $\varphi_0=0$ at $z=0$,  one can approximate the heavy quarks potential function Fig.\,\ref{Fig:VPhiHisomu=0T=0}B (an olive dashed line) by  
\be \label{VH0}
\cV_{approx,0}(\varphi)=\sum_{i=0}^{18} \mathfrak{b}_i  \varphi^i,
\ee
and by fitting the points of the graph on the domain $-13.5\leq\varphi\leq 13.5$, we obtain the following coefficients  that  are given in  Table\,\ref{table2} in  Appendix\,\ref{approx1}. It is important to note that our valid domain in these approximations for $z$ is $-8.5\leq z\leq 8.5$.

\subsection{Reconstruction of the gauge kinetic function $\ff_0(\varphi)$}

\subsubsection{the light quarks model}

The gauge kinetic function for the light quarks model takes the form \cite{Li:2017tdz}
\be
f_0(z)=e^{-c z^2+a \log \left(b z^2+1\right)},
\ee
 where the parameters $a$, $b$ and $c$ are introduced in \eqref{wfLc} and \eqref{wfL}.
Using the explicit form of the solution \eqref{Lphiz}  in  Fig.\,\ref{Fig:gkfLisoT=0}A we plot $\ff_0(\varphi)$. This function can be approximated as  
\be\label{fapL}
\ff_{0,approx}(\varphi)=\frac{1}{m_3\sqrt{2 \pi } }\exp\left[-\frac{1}{2}\left(\frac{\varphi -\varphi_0}{m_1}\right)^2\right]+m_2,
\ee
where the coefficients are the following: $m_1 =2.70895$, $m_2=8.26922*10^{-6}$ and $m_3=0.39931$.

\subsubsection{ the heavy quarks model}

\begin{figure}[h!]
  \centering
 \includegraphics[scale=0.28] {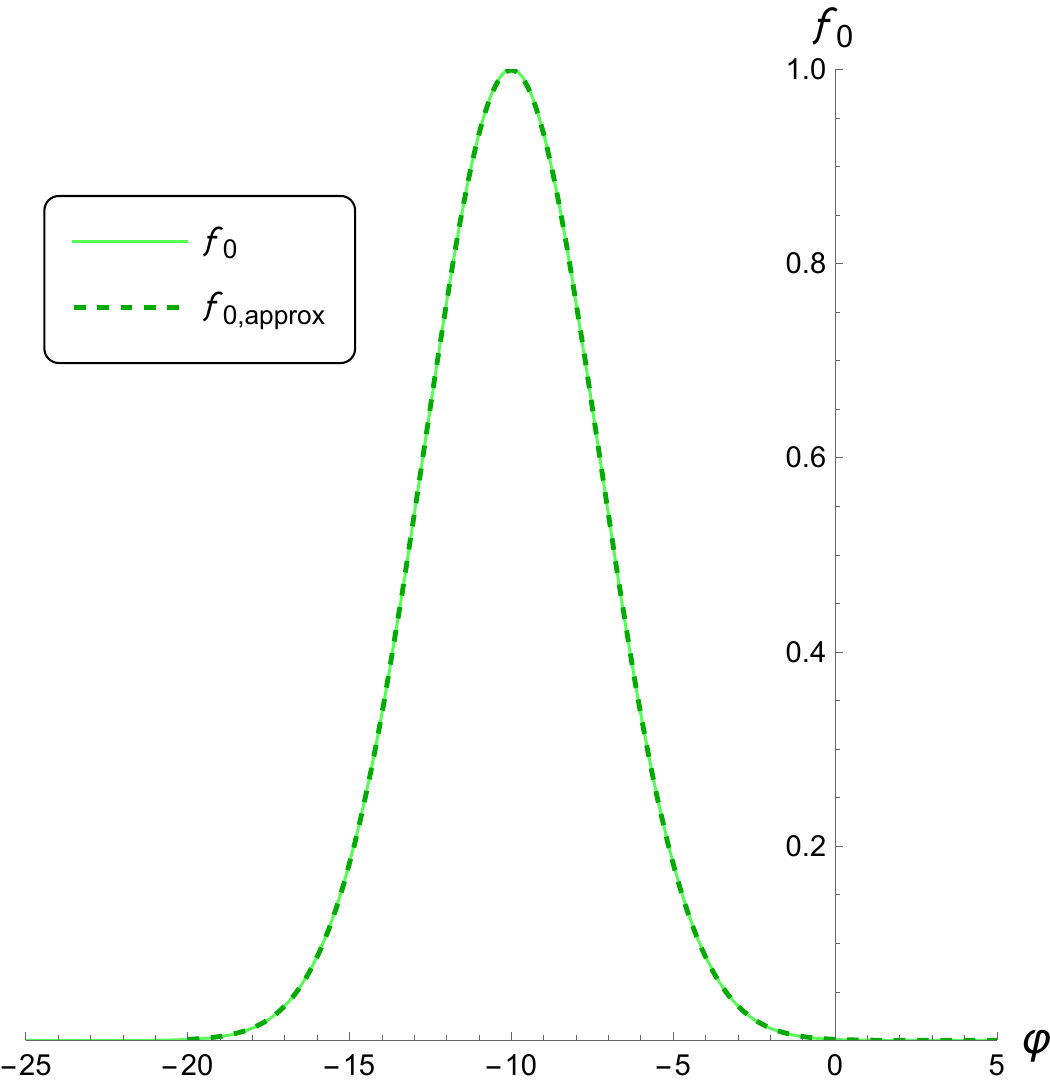} 
  \quad 
  \includegraphics[scale=0.36]{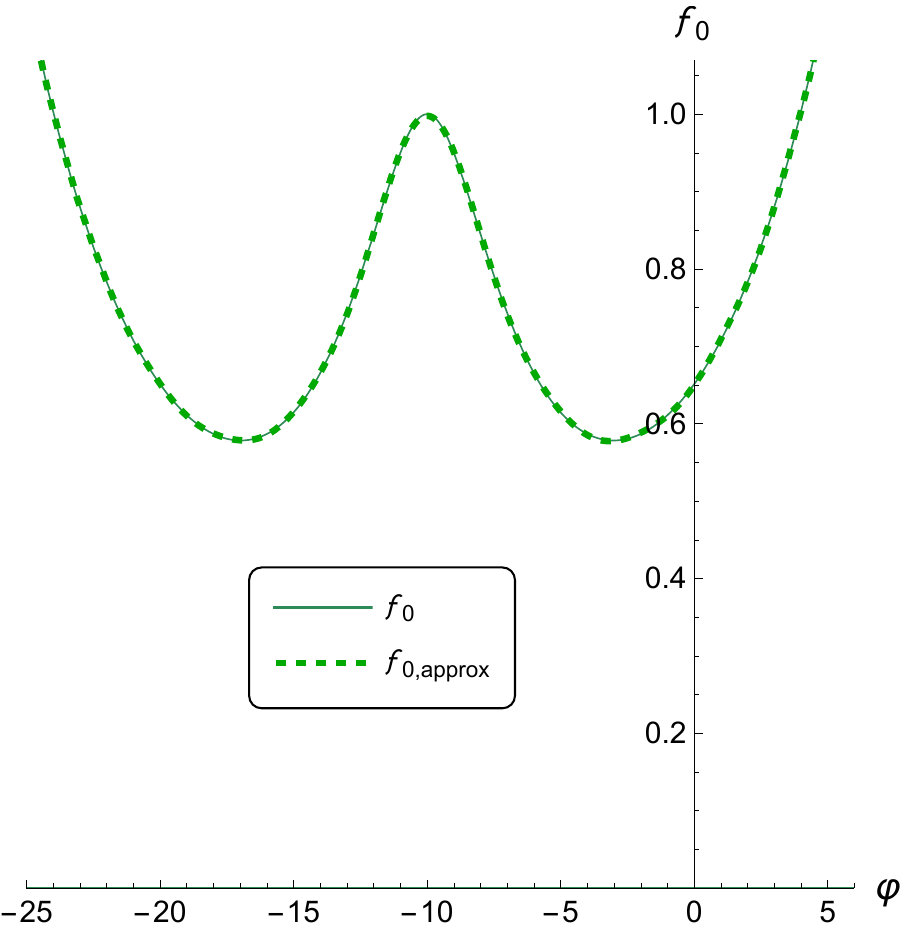}\\
  A\hspace{15em} B
 \caption{The gauge kinetic function $\ff_0$ for the light quarks (A)  and for the heavy quarks (B) as a function of $\varphi$ produced for the boundary condition $\varphi_0=-10$. The dashed lines 
 correspond to the approximation functions (\ref{fapL}) and (\ref{fapH}) for the light and heavy quarks, respectively.
}
  \label{Fig:gkfLisoT=0}
\end{figure}

The gauge kinetic function for the heavy quarks takes the form \cite{Arefeva:2023jjh,Yang:2015aia}
\be
f_{0}(z)=e^{-\frac{2}{3} \fc\, z^2+ p \, z^4},
\ee 
where $s$ and $p$ are the same parameters introduced in  \eqref{scaleHQ}. 
We utilized the form of the solution \eqref{phiHQ} to plot  
 $\ff_{0}(\varphi)$ in Fig.\,\ref{Fig:gkfLisoT=0}B. 
When the boundary condition for the dilaton field is 
selected as $\varphi_0=-10$ at $z=0$, this function can be approximated by 
\be \label{fapH}
\ff_{0,approx}(\varphi)=\sum_{i=0}^{30} w_i \,\varphi^i,
\ee
and by fitting the points of the graph on the domain $-25\leq \varphi \leq 6$, we obtain the following coefficients that  are represented in  Table\,\ref{table3} Appendix\,\ref{approx1}.\\

\section{Approximation coefficients of the potential and the gauge kinetic function } \label{approx1}

The coefficients of the approximation of the light and heavy quarks potential functions with different boundary condition, i.e. $\varphi_0=-10$ and $\varphi_0=0$ are given in the Tables\,\ref{table1} and \ref{table2}, respectively.

\begin{table}[h!]
  \centering 
  {\scriptsize\begin{tabular}{||l||c||l||c||}
    \hline
    \,\ \qquad $\varphi_0=-10$  & \,\ $\varphi_0=0$  \\ \hline
    $c_0=-8333.49$    & $b_0 = -11.996$ \\ \hline
    $c_1=-6769.87$    & $b_1 =-2.86057*10^{-12}$ \\ \hline
    $c_2=-2734.34$    & $b_2= -1.50457$ \\ \hline
    $c_3= -731.531$   & $b_3= 2.544*10^{-14}$  \\ \hline
    $c_4= -145.923$   & $b_4= -0.082475$ \\ \hline
    $c_5 = -23.1662$  & $b_5=4.19564*10^{-15}$ \\ \hline
    $c_6 = -3.05115$  & $b_6=-0.0020461$ \\ \hline
    $c_7 = -0.343354$  & $b_7=-1.87468*10^{-16}$ \\ \hline
    $c_8= -0.0337693$  & $b_8=-0.0000216737$ \\ \hline
    $c_9= -0.00294892$   & $b_9=3.45919*10^{-18}$  \\ \hline
    $c_{10}= -0.000229452$ &$b_{10}=-2.21701*10^{-7}$ \\ \hline
    $c_{11}=-0.0000156801$ &$b_{11}= -3.4441 *10^{-20}$  \\ \hline
    $c_{12}=-9.12457*10^{-7}$  & $b_{12}=-2.48019*10^{-10}$ \\ \hline
    $c_{13}= -4.34389*10^{-8}$  & $b_{13}=1.91673*10^{-22}$ \\ \hline
    $c_{14}= -1.61884*10^{-9}$  & $b_{14}=-7.31265*10^{-12}$ \\ \hline
    $c_{15}= -4.4908*10^{-11}$   & $b_{15}=-5.53005*10^{-25}$ \\ \hline
    $c_{16}= -8.662*10^{-13}$  & $b_{16}= 1.20875*10^{-14}$ \\ \hline
    $c_{17}= -1.03328*10^{-14}$  & $b_{17}=6.33555*10^{-28}$ \\ \hline
$c_{18}= -5.74044*10^{-17}$  & $b_{18}= -5.74045*10^{-17}$ \\ \hline
  \end{tabular} }
  \caption{The coefficients of the approximation for the light quarks potential function with two boundary conditions, i.e. $\varphi_0=-10$ and $\varphi_0=0$. } \label{table1}
\end{table}

\begin{table}[t!]
  \centering
 {\scriptsize \begin{tabular}{||l||c||l||c||}
    \hline
    \,\ \qquad $\varphi_0=-10$  & \,\ $\varphi_0=0$  \\ \hline
    $\mathfrak{c}_0=-22808.2$    & $\mathfrak{b}_0 = -11.9744$ \\ \hline
    $\mathfrak{c}_1=-21257.3$    & $\mathfrak{b}_1 =-2.20242*10^{-12}$ \\ \hline
    $\mathfrak{c}_2=-9781.21$    & $\mathfrak{b}_2= -1.52795$ \\ \hline
    $\mathfrak{c}_3= -2964.73$   & $\mathfrak{b}_3= -4.47153*10^{-13}$  \\ \hline
    $\mathfrak{c}_4= -666.91$    & $\mathfrak{b}_4= -0.0811821$ \\ \hline
    $\mathfrak{c}_5 = -118.904$  & $\mathfrak{b}_5= 4.56467*10^{-14}$ \\ \hline
    $\mathfrak{c}_6 = -17.5253$  & $\mathfrak{b}_6=-0.00460869$ \\ \hline
    $\mathfrak{c}_7 = -2.20197$  & $\mathfrak{b}_7=-1.74355*10^{-15}$ \\ \hline
    $\mathfrak{c}_8= -0.241625$   & $\mathfrak{b}_8=-0.0000276665$ \\ \hline
    $\mathfrak{c}_9= -0.0235074$   & $\mathfrak{b}_9=3.37277*10^{-17}$  \\ \hline
    $\mathfrak{c}_{10}= -0.00202414$ &$\mathfrak{b}_{10}=-1.26397*10^{-6}$ \\ \hline
    $\mathfrak{c}_{11}=-0.00015091$  &$\mathfrak{b}_{11}= -3.58649 *10^{-19}$  \\ \hline
    $\mathfrak{c}_{12}=-9.40185*10^{-6}$  & $\mathfrak{b}_{12}= 4.41746*10^{-9}$ \\ \hline
    $\mathfrak{c}_{13}= -4.70369*10^{-7}$  & $\mathfrak{b}_{13}= 2.13034*10^{-21}$ \\ \hline
    $\mathfrak{c}_{14}= -1.81389*10^{-8}$  & $\mathfrak{b}_{14}=-7.61899*10^{-11}$ \\ \hline
    $\mathfrak{c}_{15}= -5.14567*10^{-10}$   & $\mathfrak{b}_{15}=-6.64785*10^{-24}$ \\ \hline
    $\mathfrak{c}_{16}= -1.00593*10^{-11}$  & $\mathfrak{b}_{16}= 2.05593*10^{-10}$ \\ \hline
    $\mathfrak{c}_{17}= -1.20764*10^{-13}$  & $\mathfrak{b}_{17}= 8.49863*10^{-27}$ \\ \hline
    $\mathfrak{c}_{18}= -6.7091*10^{-16}$  & $\mathfrak{b}_{18}= -6.70907*10^{-16}$ \\ \hline
  \end{tabular} }
  \caption{The coefficients of the approximation for the heavy quarks potential function with two boundary conditions, i.e. $\varphi_0=-10$ and $\varphi_0=0$. } \label{table2}
\end{table}

The coefficients of the approximation of the heavy quarks gauge kinetic function with the boundary condition $\varphi_0=-10$ is given in Table\,\ref{table3}.

\begin{table}[t!]
  \centering
 {\scriptsize \begin{tabular}{||l||c||l||c||}
    \hline
    \multicolumn{2} { || c || }{$\varphi_0=-10$ }\\
 \hline
    $w_0= 0.649979$    &  $w_{16}= 1.31403*10^{-12}$  \\ \hline
    $w_1= 0.0505165$    & $w_{17}= 6.87886*10^{-14}$ \\ \hline
    $w_2= 0.0106962$    & $w_{18}= 7.47132*10^{-16}$ \\ \hline
    $w_3= -0.00133661$   &  $w_{19}= -6.25931*10^{-17}$ \\ \hline
    $w_4= -0.00089918$   & $w_{20}= -8.64631*10^{-19}$ \\ \hline
    $w_5= 0.000254561$  & $w_{21}= 1.02579*10^{-19}$ \\ \hline
    $w_6= 0.000146991$  & $w_{22}= 1.07317*10^{-21}$ \\ \hline
    $w_7= -0.0000111608$  & $w_{23}= -1.69210*10^{-22}$ \\ \hline
    $w_8= -0.0000108994$  & $w_{24}= -1.05002*10^{-24}$ \\ \hline
    $w_9= -4.23684*10^{-7}$   & $w_{25}= 2.89729*10^{-25}$  \\ \hline
    $w_{10}= 3.80199*10^{-7}$  & $w_{26}= 2.86221*10^{-27}$ \\ \hline
    $w_{11}= 4.82338*10^{-8}$   & $w_{27}= -4.96868*10^{-28}$ \\ \hline
    $w_{12}= -4.28331*10^{-9}$  & $w_{28}= -2.10423*10^{-29}$ \\ \hline
    $w_{13}= -1.28932*10^{-9}$  & $w_{29}= -3.46181*10^{-31}$ \\ \hline
    $w_{14}= -6.18659*10^{-11}$  & $w_{30}= -2.15509*10^{-33}$ \\ \hline
    $w_{15}= 8.43119*10^{-12}$   & $ $   \\ \hline
  \end{tabular} }
  \caption{The coefficients of the approximation of the gauge kinetic function for the heavy quarks with the boundary condition $\varphi_0=-10$. } \label{table3}
\end{table}

\newpage 
$$\,$$

\end{document}